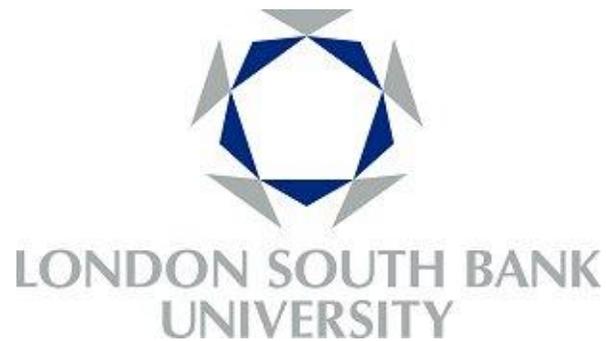

**Faculty of Engineering, Science and the Built Environment**
**MSc Project**

<u>Final Project Report</u>

# TITLE: SKIN TEMPERATURE MEASUREMENT

**Author:** SIAMAK SARJOGHIAN

**Student ID:** 2728973

**Academic Session:** 2009-2010

**Supervisor:** DR PERRY XIAO

**Course Title:** Embedded and Distributed System

**Mode of Study:** Full Time

## Table of Contents

# Acknowledgement:

This project has presented, to me, an objective, a goal, a challenge and a milestone of my final academic year in the engineering department at South Bank University.

This project marks the final hurdle that I tackle, of hopefully what would be one of the many academic challenges I have taken upon and am yet to take. However, I could not have made it without the patient support and guidance from the following people.

Firstly I want to take this opportunity to have special Thanks to My supervisor Dr Perry Xiao who helped me throughout this project by providing valuable guidance and advice as well as acquiring all components needed for this project to become a success.

Secondly I am ever so grateful to Godfrey, the lab technician for providing me full access of Lab and all components used in this project.

And finally thanks to my family and friends who have encouraged and spurred me on through this project.

# Abstract:

This report represents the design and implementation of a skin temperature measurement system. The system aims to measure the skin temperature from a sensor and send it to the PC using a USB cable to display on screen. The data needs to be updated every second.

The PIC18F4550 microcontroller has been used in this project to obtain data from the sensor and send it to the PC using USB 2.0 that has been built into the microcontroller. The microcontroller has a 10-bit Analog Digital Converting accuracy that is one of the important criteria for this design as it is going to be used for medical purposes.
As the project concentrates more on designing software than hardware, the EasyPIC4 development board was used which comes with all hardware required for this project.
The Michel Jackson diagram method was used to design and implement the coding program for the microcontroller software part of the system.
The MikroC IDE has been used to compile and load the program into PIC18F4550 microcontroller. The program for the microcontroller uses C language that aims to keep the USB link alive by using interrupt function. A sensor collects data from sensor as 4 bits and send it to the PC every second using a USB cable.

The data received from sensor, will be sent by microcontroller to the PC. The Visual Basic software was used in the PC side of device to catch and output the data on the screen. A template file for the Visual Basic program was generated by Easy HID wizard to make software programming part easier for designer. The template file was modified in such way that can operate following activities for a user purpose:
- ✓ A message that indicates if USB connector is plugged or unplugged from the PC.
- ✓ Different techniques to display data on screen (textbox & progress bar).
- ✓  Button control key to end and exit from program.
- ✓ Finally labeling the Form as well as display and control for user attention.

The USBTrace analyzer has been used in the scenario any problems occur during or after the design and construction of the software. The software enables a user to monitor the data on the USB bus that sends the data to the PC from microcontroller.

# CHAPTERS:

# 1) Introduction:

There are different techniques used to send a data from a sensor to a PC for further operation and outputting results. A requirement of the project is that the customer needs an easy to use serial cable for transferring the data from a skin temperature sensor to the host computer.

Any data communication from peripheral device to a host computer by a cable can be divided into two main groups serial and parallel data transfer. In a parallel communication, group of data will transfer simultaneously on more than one communication lines while in a serial communication the data transfers one bit at a time on single communication line. This means the ability of data transferring at each given time in the parallel communication is greater than the serial communication which makes the device faster.

There are a number of advantages of using a serial port over a parallel port such as:

- The serial ports are smaller as they have a single line to transfer therefore fewer pins compare with the parallel port needed.
- Less EMI and interference compare with the parallel port which has many wires switching over each other all the time.
- Less time for error correction compare with the parallel ports that have many lines to transfer the synchronized data.
- Finally the serial port is easy to connect compare with parallel port.

Universal Serial Bus (USB) is an external serial interface to transfer data from one electronics device to another using a USB cable.

The USB is a set of connectivity specification was developed early 1996 by Intel to establish communication between a peripheral device and a host computer which uses a serial stream. The USB cable provides a high speed, easy connectivity and automatic configuration when it is plug in. The first USB specification was developed was a USB 1.1 that had up to 12Mbps data transfer rate. However in April 2000 the USB 2.0 specification was released that had speed of up to 480Mbps which was 40 times greater than the USB 1.1.

In this project the USB 2.0 has been used to transfer data from the peripheral devices (temperature sensor device) to the host computer by USB cable.

There are a number of advantages that make companies to design their product to use the USB 2.0 instead of a serial or parallel port, more details of the USB 2.0 and some of its advantages are mentioned in following:

- Easy to use for the user as it has a plug and play function that means the user can connect any device to the PC without rebooting the system.
- Fast enough to avoid a communication bottleneck, latest standard by up to 480Mbit/sec data transfer rate.
- A standard PCs comes with 2 or 4 USB ports fit in, it will enable a user to connect up to 127 devices to the PC using the USB hubs.
- The size of a USB port is quit smaller and it is easy to connect compared to other types.
- Using the USB 2.0 gives more control to the user to control the device using its PC as it has two ways communication.
- Low cost and power consumption from the PC.

Is it not sounding old? The answer is obviously 'yes'. Technology is always progressing. It pays to move forward with new technology than try to employ the old or current. However USB 2.0 technology is widely popular and has become a standard requirement in PCs hence will be around for some time to come.

In this report convenient and up-to-date techniques and technology have been used in order to use the USB 2.0 cable for the system that enable the user to achieve their goal to satisfy their customer.

After some research over the technologies that are available in the market to convert an input data to 10 bit digital data and the chips that uses USB 2.0 technology to respond to different events on the bus, various chips were found that required a range of firmware support in order to establish USB communication.

The PIC18F4550 microcontroller was chosen to be suitable solution for this project as it satisfies the specification as it has both technologies required for this project built in (10-bit A/D conversion, USB 2.0) as well as providing more flexibility, reliability and future modification to the design.

# 2) Aim and Objectives:

## 2.1. Aim:

The aim of this project is to design a system that can measure skin temperature by an analog temperature sensor and screen out the result on a PC's screen every second.

## 2.2. Objectives:

Objectives for the system can be broken down to the following objective:

1. The primary objective of this project is to use a serial data transfer cable.
2. Secondary objective is to have good data accuracy for the system.
3. Third objective is to make more flexibility and compatibility for the system for future modification.
4. Final objective is to use an easy to understand and up to date technique to output the data on screen.

## 2.3. Specification:

1. Using the serial data transfer techniques (USB 2.0 Technology) as the customer wants it to use as an On-The-Go device.
2. Accuracy of minimum 8-bit to convert an analog input signal to corresponding digital number. As the device designed to be use for medical purposes to diagnose the skin diseases.
3. Updating the reading data from the sensor every one second. It is more accurate to get average from token results as the reading is not stable.
4. Display the input data on the screen. Use simple and easy to use technique to display the data for the user.

## 2.4. Deliverable:

The project outcome will be as follows:

- A working model of the Skin Temperature Measurement system. (Hardware)
- The software required to produce the behaviors specified. (Software)
- Research for upgrades and add-ons for the system i.e. software and hardware.
- Interim and final project documentation.
- Software and hardware testing.

# 3) Technical background and context (hardware/software):

### 3.1. Hardware:

1) Easypic4 Development Board
2) PIC18F4550

### 3.2. Software:

1) MikroC Compiler for PIC
2) PICFlash2.0 with mikroICD
3) USBTrace USB protocol analyzer
4) Microsoft Visual Basic 6.0
5) Easy HID USB wizard

# 4)   Technical Approach:

The first plan of action was to study the customer's requirements, specifications and to identify key factors of the project.

The next step was to obtain information from journals and the Internet to acquire information on how existing systems with serial ports work and to investigate what current technologies are in place that gather data from skin temperature sensors and communicate in similar fashion with a PC using USB technology.

After enough research on technologies available, the following hardware & software parts have been chosen, that are suitable for the project in order to meet the required specification set at the start of project which satisfies the criteria:

# 1. Hardware:

The hardware component of the project can break down to two main components:

## 1.1-PIC18F4550 Microcontroller:

The PIC18F4550 Microcontroller has the following features:
- ✓ The main feature of this microcontroller is that it has built in the USB 2.0 interface.
- ✓ Low power consumption (nano Watt Technology)
- ✓ Enhance Flash program and large amount of RAM memory for buffering (1-Kbyte Dual Access RAM for USB)
- ✓ Has different pin layout (40 and 44 pins)
- ✓ High Performance and speed up to 12Mb/s in Full speed mode and 1.5Mb/s in low speed mode.
- ✓ Accuracy of 10-bit to convert analog signal to digital number relatively.
- ✓ Up to 13 analog to digital converter model channels with programmable accusation time.

The project uses the 40 pins chip shown in figure(1) as it was compatible with development board provided by university lab (EasyPIC4) that has all features that satisfy the project needs.

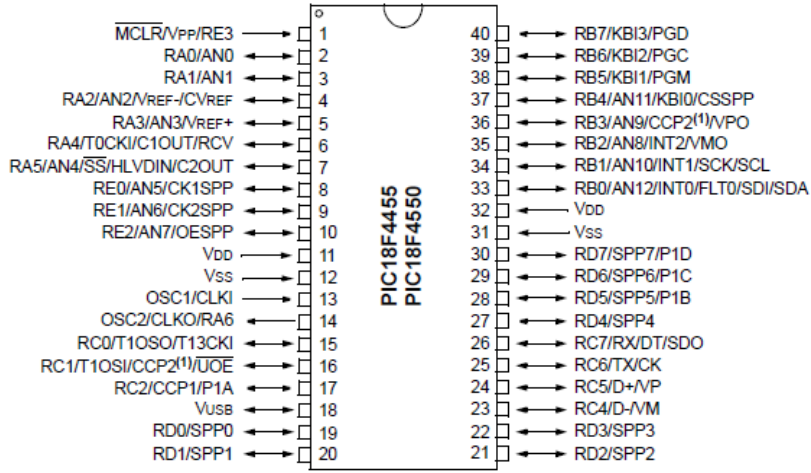

Figure (1): 40 pins PIC18F4550 Microcontroller chip

In this part of the report more details of some features and functions of PIC18F4550 that will be needed to use later on for the project will be given in following:

## USB Bus Interface:

The microcontroller supports USB interface to communicate with a host computer directly. It can transfer data with a full speed (12Mb/s) or low speed (1.5Mb/s) that allows fast communication between the device and a PC.

Figure (2) shows the infrastructure of the USB section of PIC18F4550 that use Pin23 (RC4) and Pin24 (RC5) for USB interface. RC4 is used to indicate the D- data pin and RC5 uses to indicate the D+ data pin. These pins are connected internally to two pull up resistors that can select full speed operation by connecting the resistor to the D+ data pin and low speed by simply connecting the resistor to the D- data pin.

The register UPUEN is use internally to activate and deactivate the pull up resistors. For disabling resistors simply configure UPUEN= 0x00 and vice versa to enable them the UPUEN register has to be assigned to 0xff.

These resistors can be connected externally as shown in figure but in the case of the external resistors, they can be disabled and enabled manually by connecting the resistors to 3.3 volts supply, but the user has to make sure to disable the internal resistors in the scenario of where the external resistors are to be used.

Three control registers are managing operation of the USB interface internally and 22 register are being use to manage the actual USB transaction. But MikroC language has a USB library function that can be used in order to implement the USB transaction. However there is no need to configure these registers in this project.

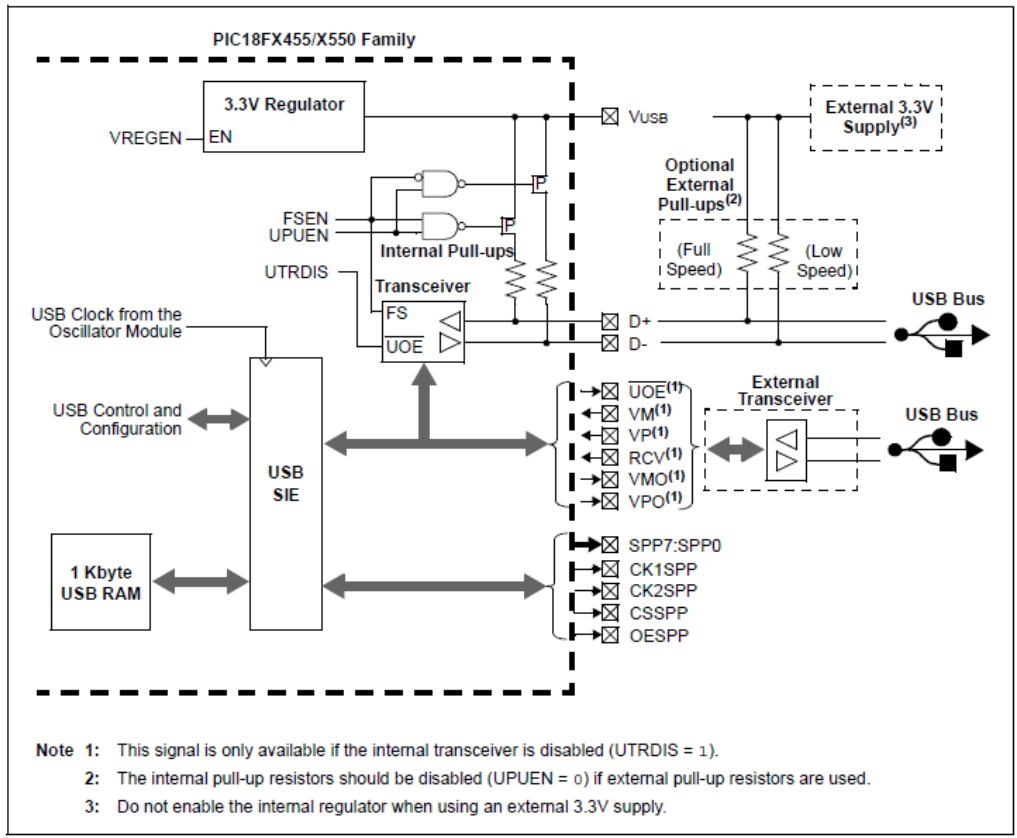

Figure (2): PIC18F4550 Microcontroller USB Overview

## 10bit analog to digital converter module:

The analog to digital converter module has 13 Input/output pins for the microcontroller with 40 pins that has been used for the project as it was mentioned before these pins provide up to 10-bit analog signal to digital number accuracy.

The module has five registers:

- ❖ A/D Result High Register (ADRESH)
- ❖ A/D Result Low Register (ADRESL)
- ❖ A/D Control Register 0 (ADCON0): To control operation of analog to digital module.
- ❖ A/D Control Register 1 (ADCON1): To control function of port pins.

- ❖ A/D Control Register 2 (ADCON2): To configure the A/D clock source, programmed accusation time and justification.

Figure (3) shows the actual register configuration in order to operate in such way which can take analog signal from channel0.

In Figure (3) the ADCON0 was assigned to zero to select channel 0 as an analog input.

| U-0 | U-0 | R/W-0 | R/W-0 | R/W-0 | R/W-0 | R/W-0 | R/W-0 |
|-----|-----|-------|-------|-------|-------|-------|-------|
| —   | —   | CHS3  | CHS2  | CHS1  | CHS0  | GO/DONE | ADON |
| bit 7 | | | | | | | bit 0 |

**Legend:**
R = Readable bit    W = Writable bit    U = Unimplemented bit, read as '0'
-n = Value at POR   '1' = Bit is set    '0' = Bit is cleared    x = Bit is unknown

bit 7-6  **Unimplemented:** Read as '0'
bit 5-2  **CHS3:CHS0:** Analog Channel Select bits
    0000 = Channel 0 (AN0)
    0001 = Channel 1 (AN1)
    0010 = Channel 2 (AN2)
    0011 = Channel 3 (AN3)
    0100 = Channel 4 (AN4)
    0101 = Channel 5 (AN5)[1,2]
    0110 = Channel 6 (AN6)[1,2]
    0111 = Channel 7 (AN7)[1,2]
    1000 = Channel 8 (AN8)
    1001 = Channel 9 (AN9)
    1010 = Channel 10 (AN10)
    1011 = Channel 11 (AN11)
    1100 = Channel 12 (AN12)
    1101 = Unimplemented[2]
    1110 = Unimplemented[2]
    1111 = Unimplemented[2]
bit 1    **GO/DONE:** A/D Conversion Status bit
    When ADON = 1:
    1 = A/D conversion in progress
    0 = A/D Idle
bit 0    **ADON:** A/D On bit
    1 = A/D converter module is enabled
    0 = A/D converter module is disabled

Figure (3): ADCON0 A/D Control Register0

In Figure (4) the ADCON1 register was assigned to zero to set all channels to analog.

Figure (4): ADCON1 A/D CONTROL REGISTER1

In Figure (5) the ADCON2 was assigned to 0xA6 to set the Ref=+5V and A/D clock = Fosc/64, 8TAD in this project.

Figure (5): ADCON2 A/D CONTROL REGISTER2

## Input/ output ports:

There are up to five ports (PORTA, B, C, D, E) are available for the PIC18F4550 microcontroller. Each port can be used as an either input or output for other peripheral features on the device.

Each port has three registers to control operation which has to initialize and activated before it can be use. The following shows the operation of these port's registers:

- ➢ TRIS register is use to specify direction of data.
- ➢ PORT register that reads the pin level of device.
- ➢ LAT register that latch the output or read and modify and write operations on the value of I/O pin.

It is better to select the PORTA as an analog input as several of its pins multiplexed with an analog inputs. PORTA is an 8-bit wide and bidirectional. To set PORTA as an input pins, the TRISA has to be set to 1. PORTB is an 8-bit wide and bidirectional which is going to be use as an output port by setting TRISB to 0.

## Interrupts:

In general interrupt is a signal that indicates attention or change in execution in programming. In this case the processor saves its states and check interrupt vector address that give the address of Interrupt Service Routine (ISR), at this point the processor will execute the ISR where the instruction for this interrupt has been stored. After the processor finishes executing of the ISR instructions, it will go back to the main program and continue from where it is left of by checking its state.

The PIC18F4550 microcontroller has two interrupt sources and a priority feature that allow each of the interrupt sources to get assigned to high priority that is located in 000008 interrupt vectors and low priority that is located in 000018 interrupt vectors.

From the name of priority can recognize that each high priority interrupt should be able to interrupt the low priority interrupt which is in the progress.

The 10 registers used to control interrupt operation has been listed as follows:

- ✓ RCON
- ✓ INTCON
- ✓ INTCON2
- ✓ INTCON3
- ✓ PIR1, PIR2
- ✓ PIE1, PIE2
- ✓ IPR1, IPR2

However there is no need to set all register for any interrupt operation. Three bits are controlling interrupt source operation:

1. Flag bit: This bit is indicated when interrupt events have occurred.
2. Enable bit: This bit allows the program that is being executing to link to the vector address in the case of flag bit is set.
3. Priority bit: This bit use to select a high priority from a low priority.

As the project contains of the interrupt operation to keep the USB function alive, there is no need to set some of the registers as well as the interrupt bits. And Timer0 function will be use as an overflow function to produce the 0.832ms delay we need before sending the keep "alive" messages.

In this part there is a need to set the registers that we need in order to disable and enable interrupt function. Before any interrupt operation is to be used it is better to disable all the interrupts in the system and this can be done by setting the Figure (6) register:

|  | R/W-0 | R/W-0 | R/W-0 | R/W-0 | R/W-0 | R/W-0 | R/W-0 | R/W-x |
|---|---|---|---|---|---|---|---|---|
|  | GIE/GIEH | PEIE/GIEL | TMR0IE | INT0IE | RBIE | TMR0IF | INT0IF | RBIF[(1)] |
|  | bit 7 |  |  |  |  |  |  | bit 0 |

Legend:
R = Readable bit    W = Writable bit    U = Unimplemented bit, read as '0'
-n = Value at POR   '1' = Bit is set    '0' = Bit is cleared    x = Bit is unknown

bit 7   **GIE/GIEH**: Global Interrupt Enable bit
        When IPEN = 0:
        1 = Enables all unmasked interrupts
        0 = Disables all interrupts
        When IPEN = 1:
        1 = Enables all high priority interrupts
        0 = Disables all high priority interrupts

bit 6   **PEIE/GIEL**: Peripheral Interrupt Enable bit
        When IPEN = 0:
        1 = Enables all unmasked peripheral interrupts
        0 = Disables all peripheral interrupts
        When IPEN = 1:
        1 = Enables all low priority peripheral interrupts
        0 = Disables all low priority peripheral interrupts

bit 5   **TMR0IE**: TMR0 Overflow Interrupt Enable bit
        1 = Enables the TMR0 overflow interrupt
        0 = Disables the TMR0 overflow interrupt

bit 4   **INT0IE**: INT0 External Interrupt Enable bit
        1 = Enables the INT0 external interrupt
        0 = Disables the INT0 external interrupt

bit 3   **RBIE**: RB Port Change Interrupt Enable bit
        1 = Enables the RB port change interrupt
        0 = Disables the RB port change interrupt

bit 2   **TMR0IF**: TMR0 Overflow Interrupt Flag bit
        1 = TMR0 register has overflowed (must be cleared in software)
        0 = TMR0 register did not overflow

bit 1   **INT0IF**: INT0 External Interrupt Flag bit
        1 = The INT0 external interrupt occurred (must be cleared in software)
        0 = The INT0 external interrupt did not occur

bit 0   **RBIF**: RB Port Change Interrupt Flag bit[(1)]
        1 = At least one of the RB7:RB4 pins changed state (must be cleared in software)
        0 = None of the RB7:RB4 pins have changed state

Figure (6): INTCON INTERRUPT CONTROL REGISTER

By setting INTCON = 0X00 we can disable all the interrupts as shown in Figure (6). The next step is to set INTCON2 = 0xF5 that set Time0 overflow and RB port as high priority and disable all the PORTB pull ups and set all the interrupt on the rising edge.

After that the INTCON3 has to be set to 0xC0 that enable the interrupt and set it as a high priority. The rest of register can be set to zero as they disable the functions.

Finally the interrupt has to be enabled where it will be used, this will be done simply by setting INTCON register to 0xE0 that enable the Timer0 overflow interrupt and all the high priority interrupts as shown in figure above.

## Timer0:

Timer0 can be using as either a timer or a counter in this project it has been used timer0 as an interrupt overflow, it is incremented on every clock by default.

Here the timer0 will be used to re-enabling the USB function every 0.832ms, this achieved by setting the following register to the calculated prescale value shown in following:

The timer0 can be set to the 256 pre-scale values and the 8 bit mode. However crystal frequency is 8MHz but the CPU is set to operate in 48MHz clock this setting will be described later.

Selecting timer value of 100 (TMR0L=100) with clock frequency of 48MHz gives the timer interrupt interval of 0.832ms that calculated below:

$T=1/f \rightarrow T=1/48M=0.02083us$

$(256-100) \times 256 \times 0.02083us = 0.832ms$

The timer0 register T0CON uses to control the all aspect. After looking at the register instruction and previous knowledge of setting, the T0CON should be set to 0x47in order to enable the timer0 and set all the bits to what it needs to be as shown below.

## Circuit Diagram:

After getting all this information about PIC18F4550, The actual hardware connection between the microcontroller and the host computer has been designed and shown in Figure (7).

The designed circuit diagram shows a sensor who is connected to channel0 (AN0) of the microcontroller. The USB cable is used to connect USB connector to the host computer. The USB connector connection was established by connecting D- to port pin RC4 and D+ to port pin RC5 of USB cable. The microcontroller clock is operating on 8MHz crystal and reference voltage was $0-5$ volts.

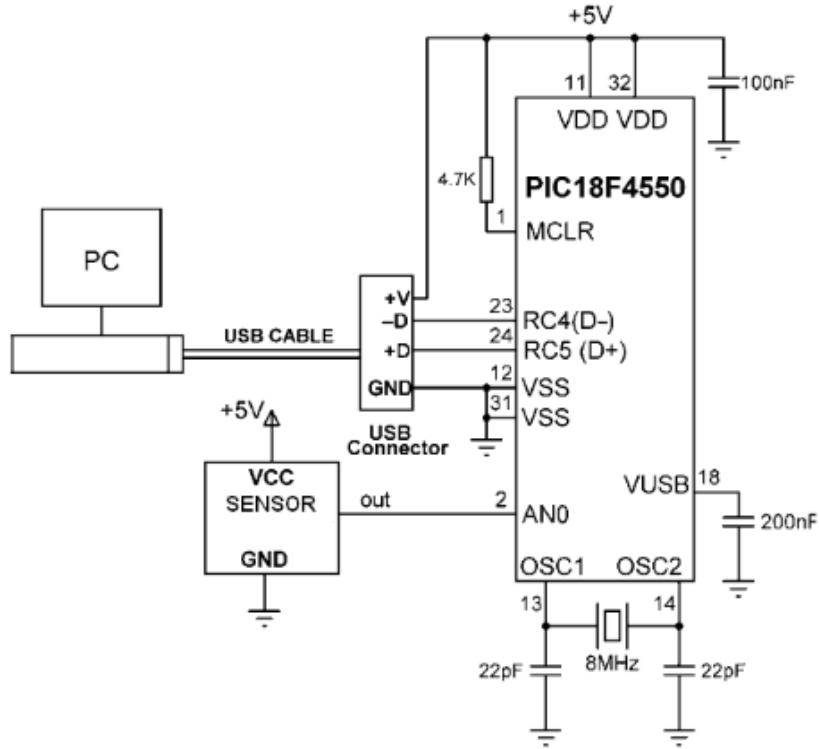

Figure (7): Circuit Diagram of Designed Project

## 1.2-EasyPIC4 Development Board:

EasyPIC Development board was designed and built by the MikroElectronica to use for the Microcontrollers with 40 pins layout. The board was recommended and provided with the university and it was familiar for me as it was used for other project previously.

The board has all the hardware features that needed for this project, therefore there is no need to design and build the hardware that leaves more time for designer to spend on the software part of the project.

In order to activate the hardware parts of development board that is needed for the project. The following setting and connectivity has to be established by the user.
In this section more detail of the hardware parts of EasyPIC4 board and how to set the parts in order to operate as expected given in Figure (8):

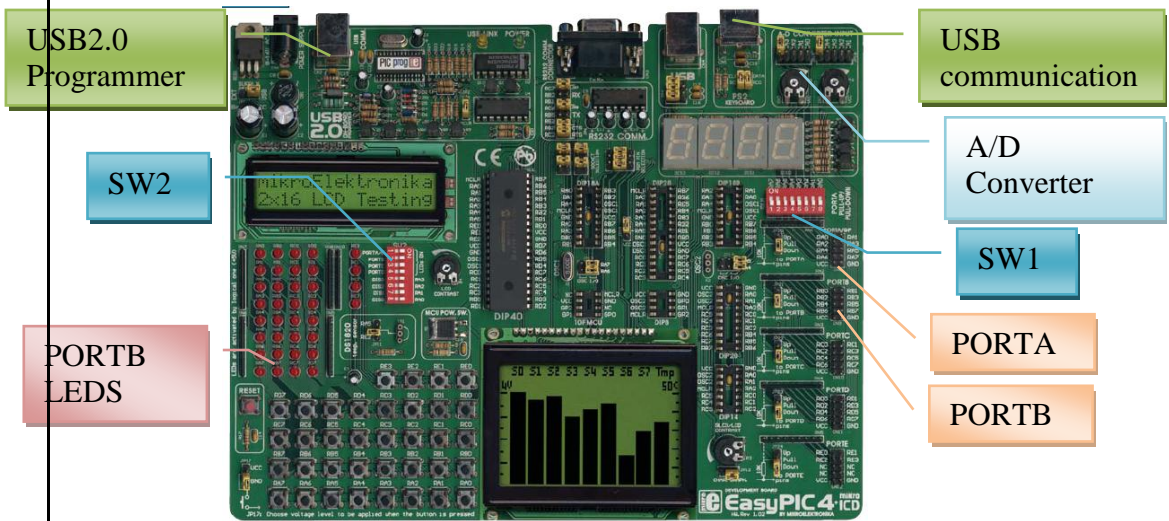

Figure (8): Easy PIC4 Development Board

The hardware components used in this project have been indicated in Figure (8) and their settings have been described in following:

## Switches:

The board has two sets of switches (SW1&SW2) that have two positions (ON&OFF) which activate and deactivate the connection. First sets of switches (SW1) are used to enable connection between the microcontroller and the input ports as well as the external pull up/down resistors. In order to avoid effect on the input port voltage level these resistor should be disabled.

In this project it is better to leave all the switches OFF as the analog input is going to be measured. The next sets of the switches (SW2) are used to enable the LEDs connected to the PORTs. In this project the PORTB is enabled by placing the switch in the ON position and leave the rest of the other ports in the OFF.

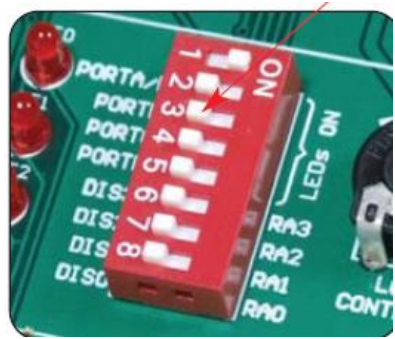

Figure (9): The board Switches

## Power Supply:

The board can be powered up by either regulated power from the USB cable or by the external power supply. In the case of using regulated power from the USB programmer cable the power will be supply while the board connected to the PC bye the cable. In this case jumper JP1 should be set to the right hand position as shown in Figure (10):

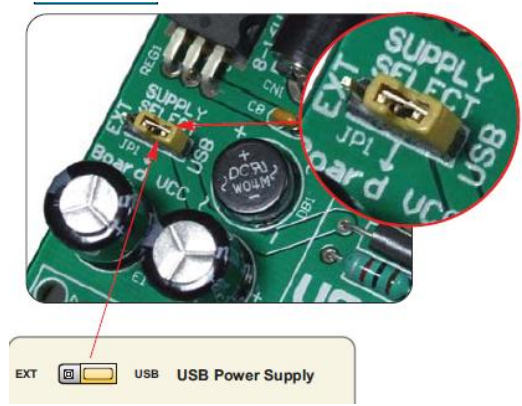

Figure (10): Supply Set Jumper

## USB 2.0 Programmer:

In order to use this port, the PICFlash2 programmer software which comes with the board or will be provided on the Microchip website which has to be downloaded and installed onto the PC.

This software enables the user to load a program into microcontroller. There is no need to reset the microcontroller as programmer will reset it automatically. There is one jumper (JP5) located for this part that should be set as a default position.

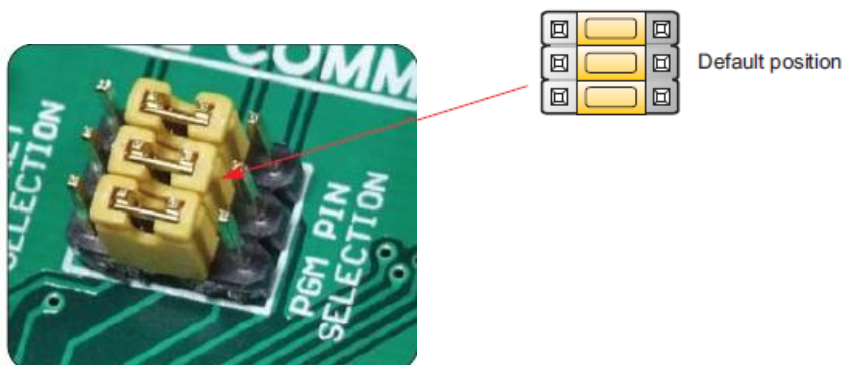

Figure (11): USB 2.0 Programmer Jumper

## MikroICD Real Time Hardware in Circuit Debugger:

MikroICD Debugger is a tool that enables the user for Real Time debugging on hardware. MikroICD programmer provides a service for the user to check the program variable values while the program running on the microcontroller using a Special Function Register (SFR) and EPPRAM. This tool can be used through MikroC compiler, it can communicate with the compiler and supports common debugger commands that shown in Figure (12).

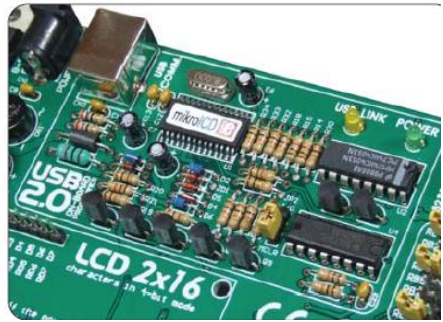

Figure (12): MikroICD Debugger

## LEDs:

Light Emitting Diodes (LEDs) will be used commonly on the board to display output. The board contains of the 36 LEDs of which 8 LEDs are used for each the PORTA, B, C, and D except PORTE that has 4 LEDs. In this project PORTB was selected as an output by changing the position of SW2 to ON position, and monitor output value all the time while the program was tested.

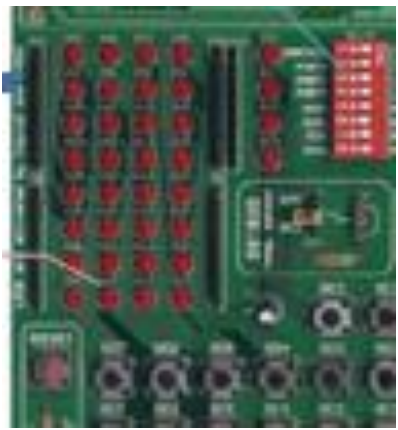

Figure (13): Board's LEDs

## USB Communication:

This USB connector can communicate with those microcontrollers that have the USB technology built in such as the PIC18F4550 microcontroller. To enable communication between the microcontroller and connector the jumper set group has to set to the right position that disconnect the pins RC3, RC4 and RC5 from the rest of the system and connect it to the USB connector.

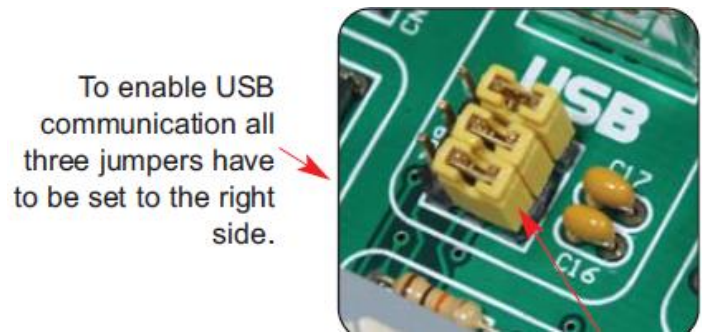

Figure (14): USB Communication Jumpers

## A/D Converter Input:

The board has two potentiometers built in which can work with an Analog Digital Converter (ADC) to produce the input range of 0 to 5 volts. These potentiometers can be used for testing purposes in the case of unknown sensor is going to be used.

In this project these potentiometers have been used as the customer did not provide any details of sensors that are going to be used as two analog signals were presented by these potentiometers in the same time.

To connect these two potentiometers the following connection has to be established. The jumpers group JP15 can enable connection between P1 and one of the following pins: RA0, RA1, RA2, RA3 and RA4.
The jumper group JP16 can enable connection between P2 and one of following pins: RA1, RA2, RA3, RA4 and RA5.
In this project the jumper JP15 has to be connected between P1 and pin RA0.

The corresponding switch (SW1) has to be in the OFF position that disconnect the Pull up/down resistor from PORTA of the circuit, this gives measurement result without any interference.

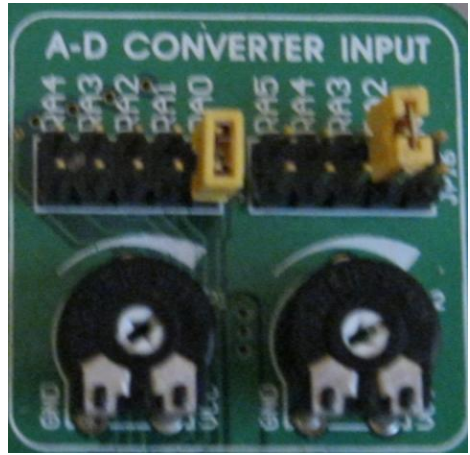

Figure (15): A/D Converter Set Jumpers

## Direct Port Access:

There are five direct ports were provided on the board to enable the user to connect to any peripheral such as sensor in this case.

Each groups of port pins are connected to one of the PORTs A, B, C, D and E which have 10 pins connecter VCC, GND and up to 8 port pins.

The user has to make sure the connecter is disabled from on board peripheral by checking the jumper next to it before connection to the external peripheral has been established.

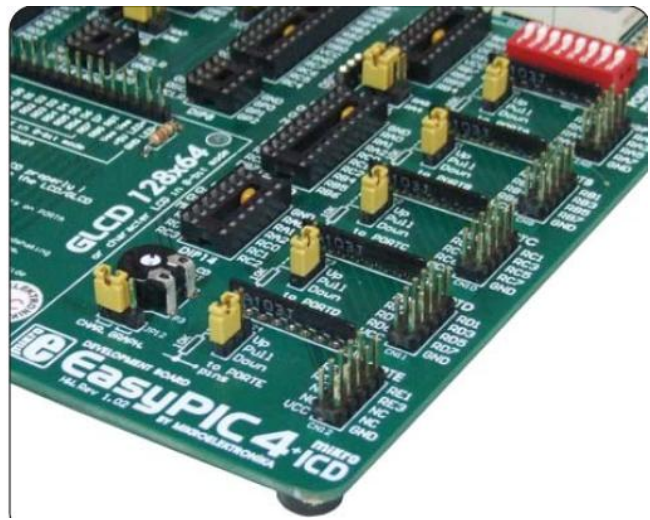

Figure (16): Direct Ports Access

# 2. Software:

The software part of this project can be broken down into two different programs. The first program has been designed and implemented was the microcontroller program, and second program was the host computer program. By using the following techniques that enable the user to develop application for this embedded system.

The project consists of two software part as it is mention before in the report:

**Microcontroller Software:** This part consists of design and program the microcontroller that can take input data from channel0 (AN0) and use USB cable to send it to the PC each second for further operation and screen out.

**PC Software:** In this part the program has to be design and implement to detect any USB connection and catch data that was send from the microcontroller and screen out for the user.

## 2.1-Microcontroller Software:

The main part of this project is consists on the software design that should satisfy the customer requirement. The hard ware part that was described in the previous part of this report gives more information of the PIC18F4550 microcontroller and function's which can be used in this part of the project.

The software program for the microcontroller has to get the data from the sensor and send it to the PC using a USB cable. However this data has to be updated every second.
The PC software part of project will be described later in this report which uses Visual Basic software to catch and output input data on screen.

Before we start to design the software program for the microcontroller, the operations that have to be done to satisfy the specification has been broken down into following Jackson Structure Programming (JSD) method:

## 2.1.1-JSD diagram:

The Jackson System Development (JSD) diagram is a liner method for software development that uses life cycle structure to produce final code for the microcontroller. Following JSD diagram designed for the microcontroller software part of the system using Microsoft Visio.

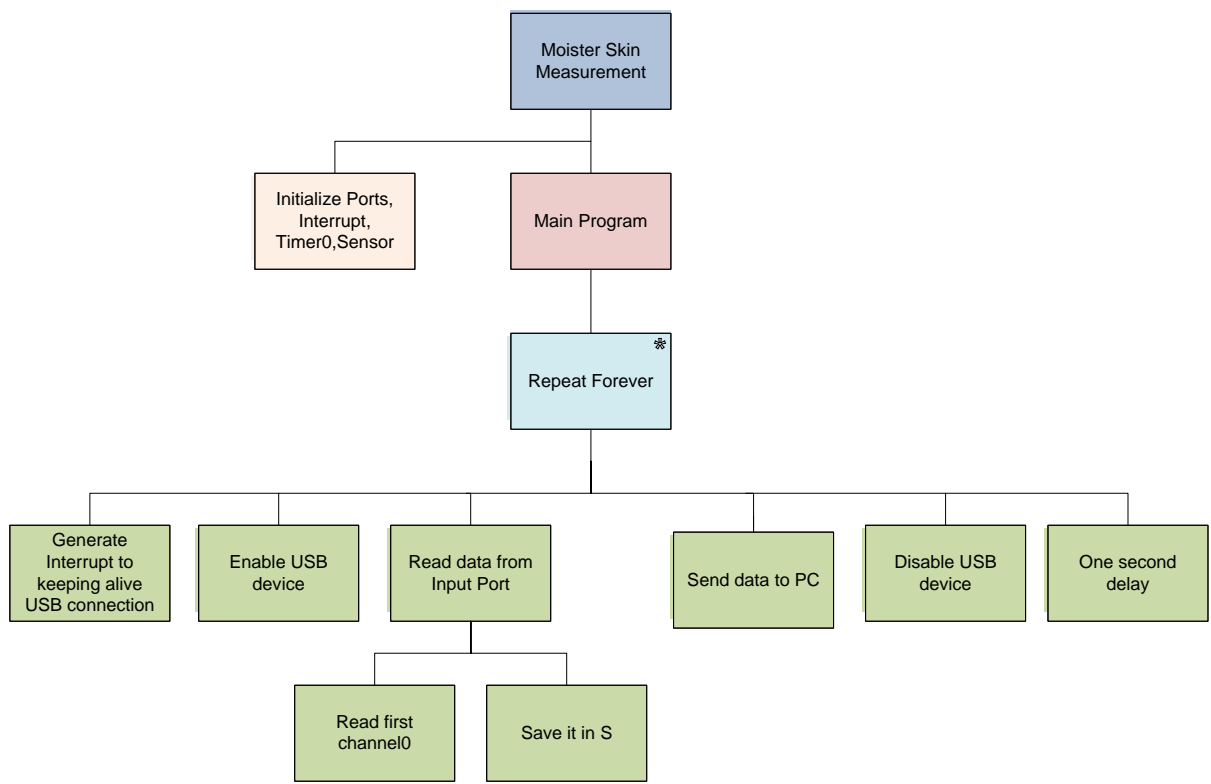

***Jackson circuit Diagram***

**Diagram explained:**

1) The first step in the diagram was initialization of the ports, variables, TIMER0 and interrupts.
2) Next, the main program is executed which is an iteration type of box that indicates an infinite loop (main loop). The loop is important because the microcontroller reboots if the end of the program is reached.
3) Generating interrupt to making the USB link alive during the operation of sending data to the PC. Enabling the interrupt and Timer0 has been described in the previous microcontroller part of this report.
4) The next step on the diagram is to enable the USB device on the chip that carries the data.
5) On the next step the data is read from the specific channel of the input port on the board and the read data was saved in the variable S.
6) In this step the data has been saved on the variable S has to be send to the PC using the USB cable.
7) Disabling the USB device is next step to make sure there is no more data can be detected.
8) And finally 1 second software delay is produced in this part of the program that makes this loop wait for one second before actually going into next loop.

Following JSD Diagram, the C program Code is written for microcontroller but before writing code it is best to understand the functionalities of the MikroC compiler that is going to be used in this project.

The reason MikroC has been chosen for the project was that as it has the HID library function built in as well as some operation such as the HID terminal can help us to interface the microcontroller with the PC to send and receive data.

## 2.1.2-MikroC Compiler for PIC:

MikroC is one of the rich feature development tools for the microcontrollers. It is provide the easiest solution without compromising the performance or control for developing embedded systems.

It provides a successful match featuring highly advanced IDE, ANSI compliant compiler, broad sets of hardware libraries, comprehensive documentation and plenty of ready to run examples.

In this part the software has been downloaded and executed by double clicks on MikroC icon on desktop which bring up the window where a new project can be created for this application.

Each application can contains of a single project file (name.ppc) and one or more source file (name.c). The compilation can be done on those file which are part of the project.

First step is to create a project that can be done easily by dropping down the menu and then project>new project which take us in another window that has following parts to fill up:

1. Project name: Each project has to have a name that can be chosen by the user in this project the name selected to be temperature.
2. Project path: this can be any location in c directory or anywhere else where the user wants it to save the file. In this project is selected as default that takes us to the Mkiroelectronika examples folder.
   C:\ProgramFiles\Mikroelektronika\mikroC\Examples\EasyPic4\extra_examples\HID-library\
3. Device: In this part the PIC to be used has to be selected. In this project PIC18F45550 was selected as it is going to be use on board.
4. Clock: This gives more capability to the user to select clocking user want to use on their project. In this project 48MHz was selected.

The rest of device flag has to be selected by default.
By clicking on save and then ok the software will create new project and empty source file that can be name by the user later.

In this file the C coding for microcontroller is written that operates as it has been explain by JSD Diagram method previously.

In this part before writing the code software for this project a brief description of some of the library functions such as the USB HID and the ADC used in this application is as follows:

i. **ADC library function:**

   This library function can be used to read 10-bit unsigned data value from the specific channel. This value is saved into unsigned long variable for more operation.
   The prototype for this function is:
   Unsigned Adc_Read (char channel);

ii. **USB HID library function:**

   MikroC has a library that enables the user to connect and work with the Human Interface Device (HID) via Universal Serial Bus (USB).
   The HID is a type of computer device that enable the user to interact and take input directly from other input humans (microcontroller input).

   In the case of using the PIC18F4550 microcontroller or any other PIC that has USB device built in the following USB library function can be used:

- **HID_Enable:** This function has been used to enable the USB communication that has to be called before any other function of this library and it does not return any data. The function has two arguments which have to be set before using them (read buffer address, write buffer address).
- **HID_Read:** This function is used to get data from the USB bus and store received data into the read buffer. This function does not have any argument but it does

> return the number of characters that has been stored in the receive buffer previously.

- **HID_Write:** This function use to send data from the write buffer to the USB bus. The function does not return any data. To use this function has to be careful to copy the same buffer name that has been specified in the initialization and length of data that is going to be send has to be specified in the function argument.

- **HID_Disable:** This function always use at the end of this library function and disable the USB communication. This function does not have any argument and returns no data.

Any project that uses the USB library should include a descriptor source file that contains of a vendor ID and name, a product ID and name, a report length (input/output) and other relevant information. In MikroC this can be created directly by using the USB HID terminal tools.

Bring down the menu and then tool>HID terminal after clicking on the HID terminal, it will be open. On this window click on the descriptor that will go to next window. In this window fill up all the descriptor information as shown in Figure (17) and click on create button.

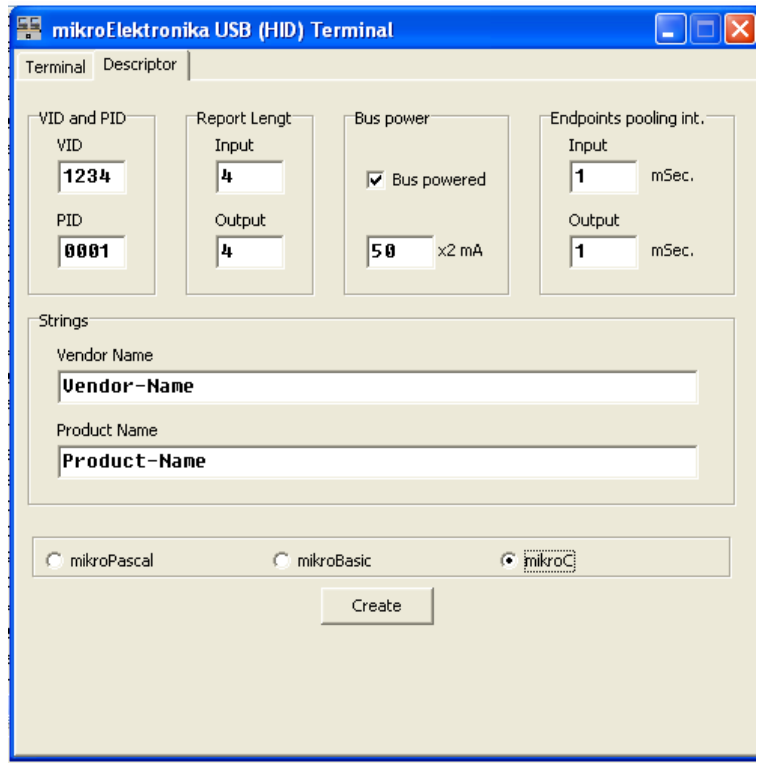

Figure (17): Mikro Electronika USB Descriptor

The vendor ID has been set to 1234 (equvalient to decimel 6460) and the Product ID to 1 which was provided by the USB-IF (www.usb.org) which is non-profit corporation that provides software and hardware to help developing and testing products.

By clicking on create button and selecting the project location the compiler will create the USBdsc.c descripter file that shown in appendix.

This file has to be added to the project with either using mikroC IDE tools or include to the project by using #include on the program file.

The program has to have the function to keeping alive the USB connection by sending message to the PC every few millisecond.

This can be achieved by setting up a interrupt that uses the Timer0 to over flow every 0.832ms and inside interrupt service routine the USB function HID_InterruptProc has to be called following that the Timer (TMR0L) is reloaded and the timer interrupts are re-enabled just before returning from the interrupt service routine.

## 2.1.3-Microcontroller clock:

The PIC18F4550 microcontroller requires a 48MHz clock frequency for its USB module and the clock range of 0 to 48MHz is needed for the microcontroller's CPU. In this project the CPU clock has been set to 48MHz. this setting has been done using MikroC compiler.

The circuit is shown in Figure (18) illustrate the PIC18F4550 clock circuit.
As it is shown in the circuit it consist of following components that has to be set in order to clock in the specific clock frequency:

- A 1:1 - 1: 12 PLL prescaler and multiplexer
- A 4:96MHz PLL
- A 1:2 - 1:6 PLL postscaler
- A 1:1 - 1:4 oscillators postscaler

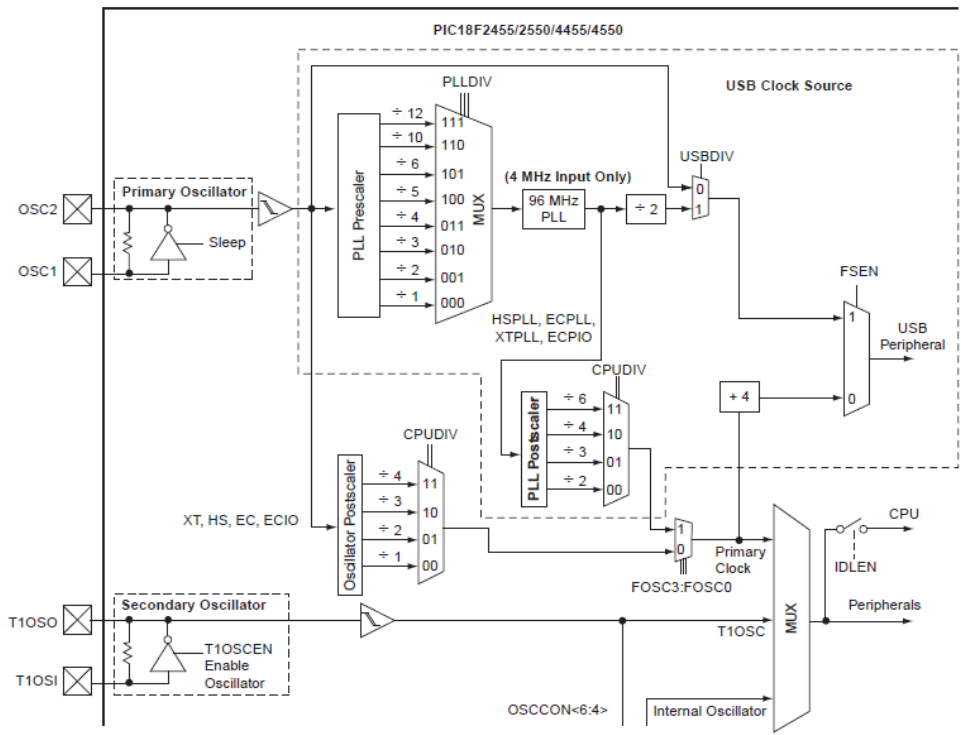

Figure (18): PIC18F4550 Microcontroller Clock Diagram

As it has been mentioned before the crystal frequency is 8MHz. In order to set the microcontroller and the USB module operate with 48MHz clock, the following has to be selected in the Edit Project option of MikroC IDE:

- The _PLL_DIV2_1L was selected that makes 8MHz clock is divided by 2 to produce 4MHz at the output of the PLL prescaler multiplexer. The output of the 4:96MHz PLL is 96MHz now. It needs further to be divided by 2 to give 48MHz at the input of multiplexer USBDIV.
- The _USBDIV_2_1L has been selected to provide a 48MHz clock to USB module and 2 for the PLL postscaler.
- The CPUDIV_OSC1_PLL2_1L has been selected to choose PLL as the clock source.
- The _FOSC_HSPLL_HS_1H has been selected to choose a 48MHz clock for the CPU.
- The CPU clock has been set to 48MHz using the Edit Project option of MikroC IDE.

Figure (19) shows the setting that has been done in the MikroC Edit Project option in order to select 48MHz clock frequency for the USB operation and the CPU clock.

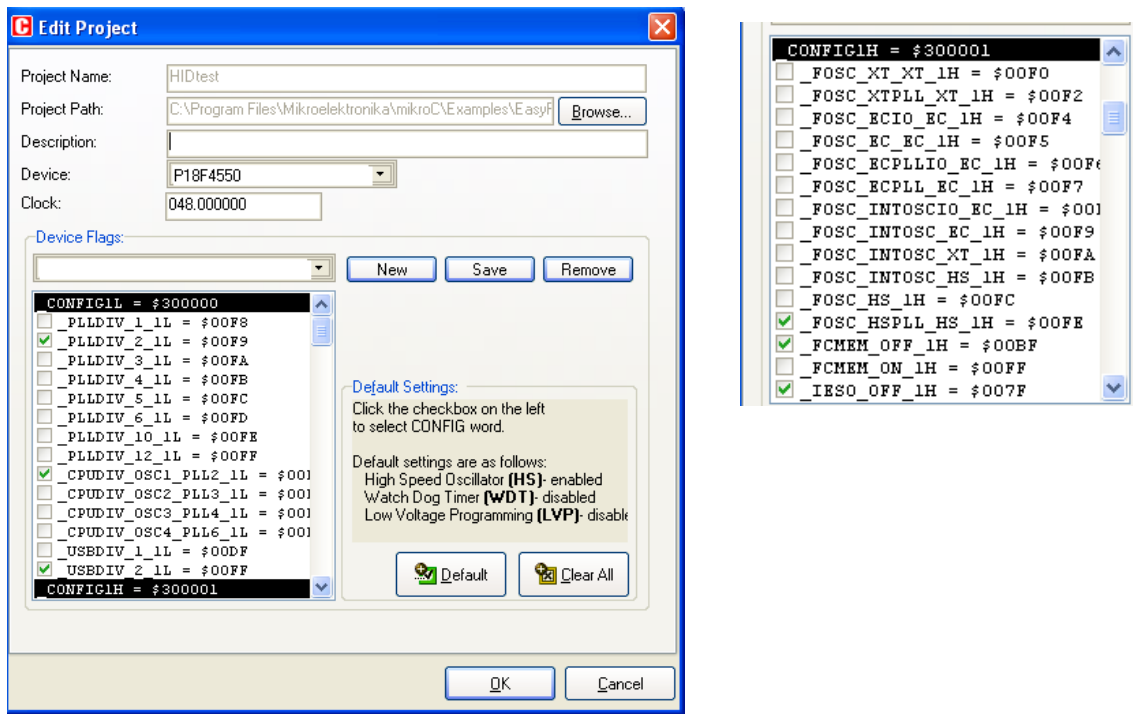

Figure (19): Edit project window

The source code main program file for the microcontroller that shown in appendix 1 has been written base on the JSD Diagram and the information was provided in previous part of the report. More details of source code will be given in the following part of report.

## 2.1.4-Main Program for Microcontroller:

This program has been written to connect the PIC18F4550 microcontroller to the PC using a USB cable. One of main important aspect that should be taken into consideration when the USB link is going to be use in any embedded design is to make sure that the USB link is alive during any operation of the bus.

The microcontroller is supposed to obtain the data from a temperature sensor that has been connected to the channel0 of the microcontroller. The data was obtained from the sensor is in Celsius and has to be sent to the PC every one second to get display on the screen.

The Visual Basic is used to display the sensor value on the screen that will be explained later in this report. The PIC18F4550 is operates from the 8MHz crystal however the CPU clock frequency and the USB module had set to 48MHz.

The temperature data sent to the PC as 4 digit integer number and it has to be in Celsius.

The main program has been broken down into following sections that make it easier to explain and understand.

1. **Include USB Descriptor:** The first and important part of this project is to include the USBdsc.c file to the main program which enables the program to use the USB library for the function that will be use later in main program file.
   This can be done in two ways either using MikroC IDE tool that enable user to add file to the project or simply added to main program using #include option follow by location of file.

2. **Define Variable:** The second part of each coding will be variable definition that is going to be used in coding. As it shows in code there are a number of unsigned variables has been used to carry the sensor data and three arrays. One to store first read data from the sensor and another two to be used for write and read buffer, to get easily spot the write buffer name has been changed to the temperature that is going to be used to carry data to the PC.

3. **Initialization function:** In this part of the code, the PORTA was set as an input by assigning the TRISA to 0xFF and the PORTB as an output by assigning the TRISB to 0x00. The input was set as an analog and +5 reference by setting ADCON1 = 0x00 and analog to digital clock was set to Fosc/64, 8TAD by setting ADCON2 = 0xA6. The way to do this setting was described with the registers well in the PIC18F4550 part of this report.

4. **Interrupt function:** The interrupt function was used in this program to keep alive the USB link during communication or sending data.
   The function is keeping link alive by sending a message every 0.832ms and is placed on interrupt service routine (HID_InterruptProc). Timer0 is loaded with 100 that produce 0.832ms overflow time delay and enabling and disabling the interrupt all was described previously on the PIC18F4550 part of the report.

5. **Removing blank function:** This function is designed to remove extra blanks and digits, as only 4 digits of input data have to be screened out on the PC's screen.

6. **Main loop:**

The initialization function (System_init) is the first function to be called in the main loop after initialization the interrupt function that contains of three functions is activated as following the first function (Interrupt_Dis) disabled all the interrupts, the second function (Timer0_init) activated timer0 as well as loading it with 100 to make 0.832ms delay and the third and last function (Interrupt_En) is to enable the interrupt. All setting part of these function has been describe in details previously in the PIC18F4550 part of this report.

The next function that has to be activated in the main loop is the HID USB. The HID enable function was described previously in the report used to enable the Human Interface Device before any other the USB functions. The function, as described before has two arguments write buffer (&Temperature) and read buffer (&Read_buffer). The write buffer was used to store and send input data to the host computer.

The next process was infinite loop which was added to the main loop as the microcontroller never stops the main program has to be in infinite loop.
The first operation is the infinite loop is to read the input voltage data from channel0 (AN0). In this design the input channel is connected to the potentiometer on the board. The input data from potentiometer will be stored in the "Vin" variable, was defined before in the initialization. The store data will be converted to voltage and get stored in the Voltage variable by following equation:

Voltage = Vin x 5/1024

Next step is to convert this input voltage to temperature in Celsius that can be done by following equation:

Celsius = (Voltage $\times$ 100)

As the data needs to screen out as a 4 digit integer, the Celsius data was converted to integer and the result was stored on the Cint variable. The following equation was used to do that:

Cint = (int) Celsius

To be able to test the software and checking the output data the PORTB was define in this part which enabled the user to output data on the PORTB LEDs for testing purposes.

As the USB function can only send data in string, the integer input data was converted to string and was stored in the op array that defined in the beginning of the coding program.

The op array can take up to 12 data sizes. As 4 digits has to screen out the remove blank function will be use in this line to take 4 first data from the op array and remove leading blank of the array. The modify op array was stored it in the temperature buffer array (write buffer) to be send by the next function. The next function is Hid_Write (&Temperature, 4) which was used to send 4 digits to the PC using the USB bus.

As the program should updated every second to satisfy specification, in this part one second software delay was added to the program. This was done by following line of code.

Delay_ms (1000);

The program is in the infinite loop to get input data from a sensor and send it to the PC via the USB bus every one second.

Finally the Hid_Disable function was used to disable the human interface device at the end of the main loop however the program will be stock in the infinite loop after that. The main program for the microcontroller was shown in appendix 1.

## Program testing:

After the program was written, the design code was checked using MikroC IDE software. The software has the HID terminal feature which enables the designer to output the data being send by the microcontroller.

The coding program was loaded into the microcontroller using the PICFlash software by connecting the USB programmer to the board and the PC. The USB cable was connected to the PC and the data was monitored from HID terminal by simply selecting device which was added when the USB cable was connected.

The data was output on the USB terminal every second as was expected. Figure (20) shows the data was output by the HID terminal.

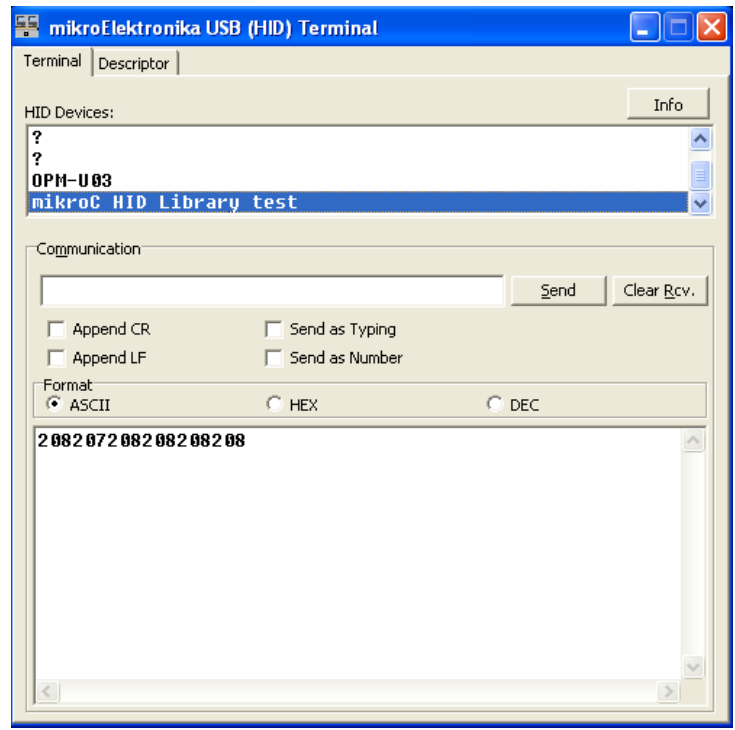

Figure (20): Mikro Electronika USB Terminal

## 2.2-PC Software:

The PC software part of this project is based on Visual Basic. The software program was written to achieve the detection of any USB connectivity operation as well as outputting the input data sent by the microcontroller on the USB bus as 4 bit integers.

The Visual Basic program was used in this project is base on the USB utility which known as Easy HID USB wizard.

The following software programs were used in the project to achieve the aim of screening out the data on the host computer's screen.

### 2.2.1-Easy HID USB wizard

Easy HID USB wizard is a software that can be use to generate an initial template file for the USB PIC and Visual Basic framework.

The software was developed by Mecanique and it can be downloading free of charge from their website (www.mecanique.co.uk).

The software was designed to work with the USB 2.0 and the user does not need to develop any driver to use it as the window XP operating system has the HID based USB driver. The generated code by Easy hid can be expand in such way that fulfils requirement.

The following step was taken in order to generate the template code for the Visual Basic:

- ➢ The first step was to load the zip file and install software by double clicking on set up icon.

- ➢ After the software was installed, it can be run by double clicking on the icon that appeared on desktop after installation.

- ➢ After running the software, the following window will pops up that needs to be filled up by company name, product name and serial number as it shown:

Figure (21): Introduction Form

- ➢ By clicking next, the next form will be display that has the Vender ID and Product ID as it shown in following form. The vender IDs are unique and issued by the USB implementer (www.usb.org). Mecanique has its own vender ID and can issue a set of Product IDs for each product with low cost to be use all over the world with the unique vender and product ID. In this project vender and product ID was selected as shown in Figure (22) which is use for testing purposes.

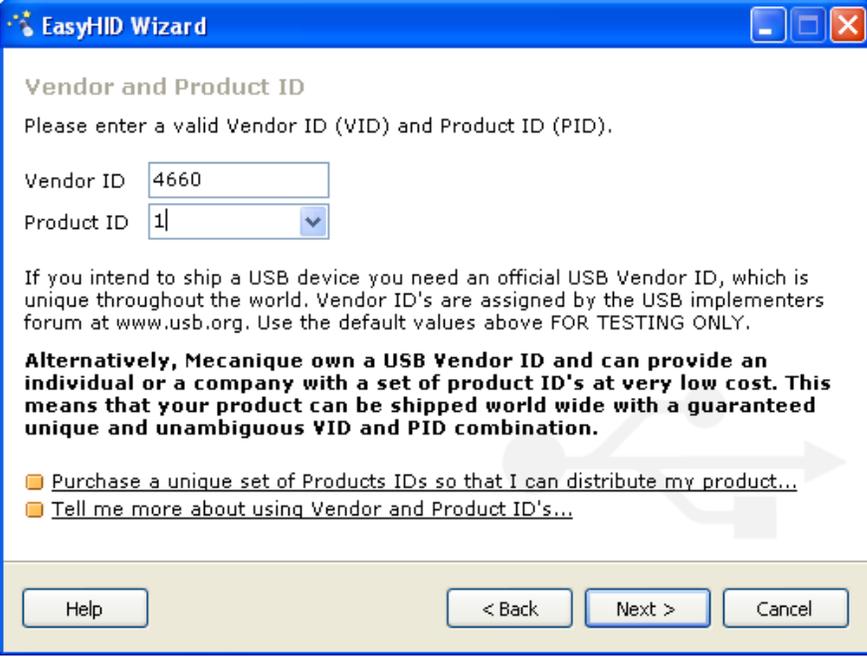

Figure (22): Vender and Product ID Form

- The next form that appear after clicking next, has different parameters such as polling interval for input and output as well as bus maximum power consumption. But the main important parameter in this part that has to be changed according to the project requirement is the input and output buffer size.

  These sizes define the number of bits will be send or receive during a USB transaction between the PC and the Microcontroller. In this project both of the input and output was selected to be 4 bits as 4 bits input data are going to be receiving from the microcontroller.

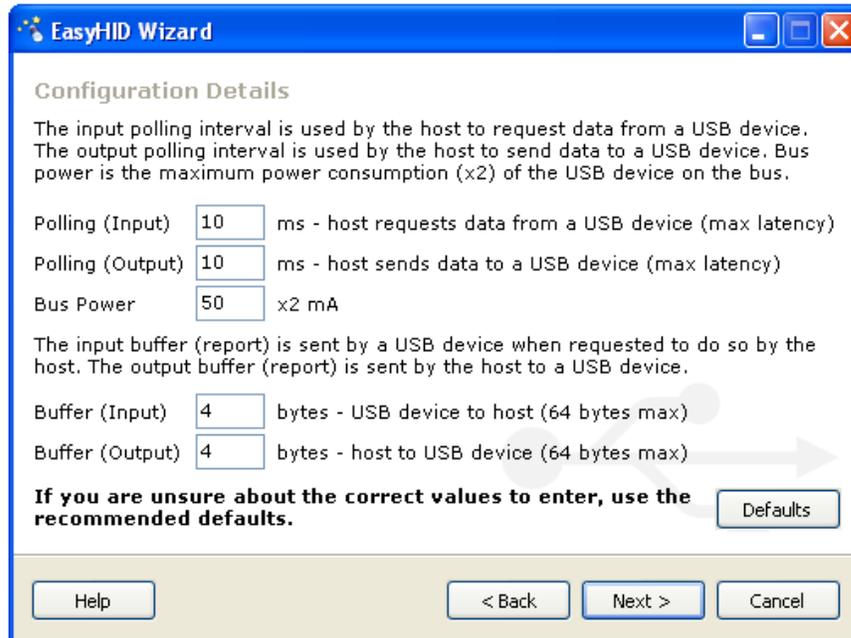

Figure (23): Configuration Form

- In the next form the project name and location has to be selected as well as compiler and Microcontroller that is was selected to be used by the project. these part and location was selected as shown in Figure (24):

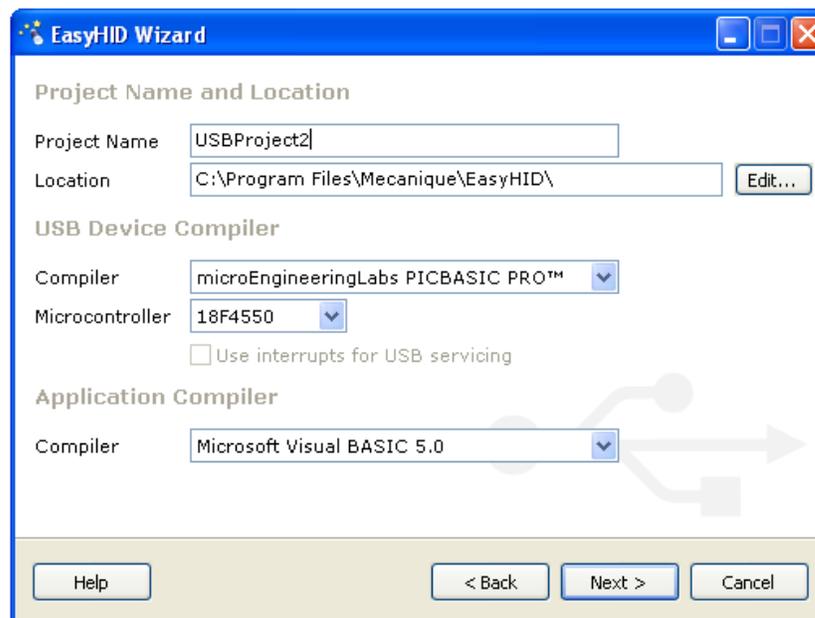

Figure (24): Project Name and Location Form

- Clicking next will generate the Visual Basic template file in the selected directory which was defined in previous form as shown in Figure (25):

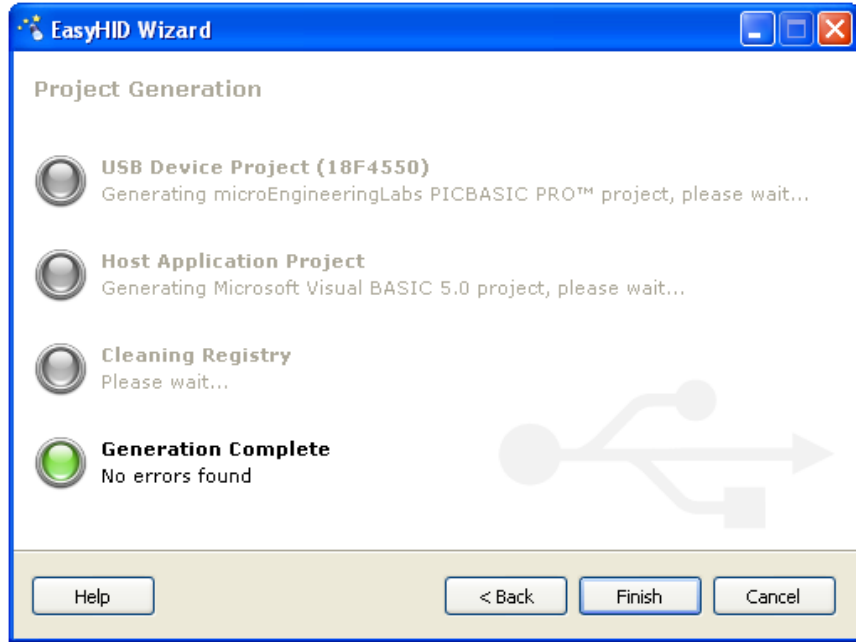

Figure (25): File Generation Form

The file was generated by the Easy HID software has consist of a blank form (FormMain.frm), a module file (mcHIDInterface.BAS), and a project file (USBProject.vbp) as shown in Figure (26). The generated project file can be opened by the Visual Basic for more expansion and modification to satisfy the project requirements.

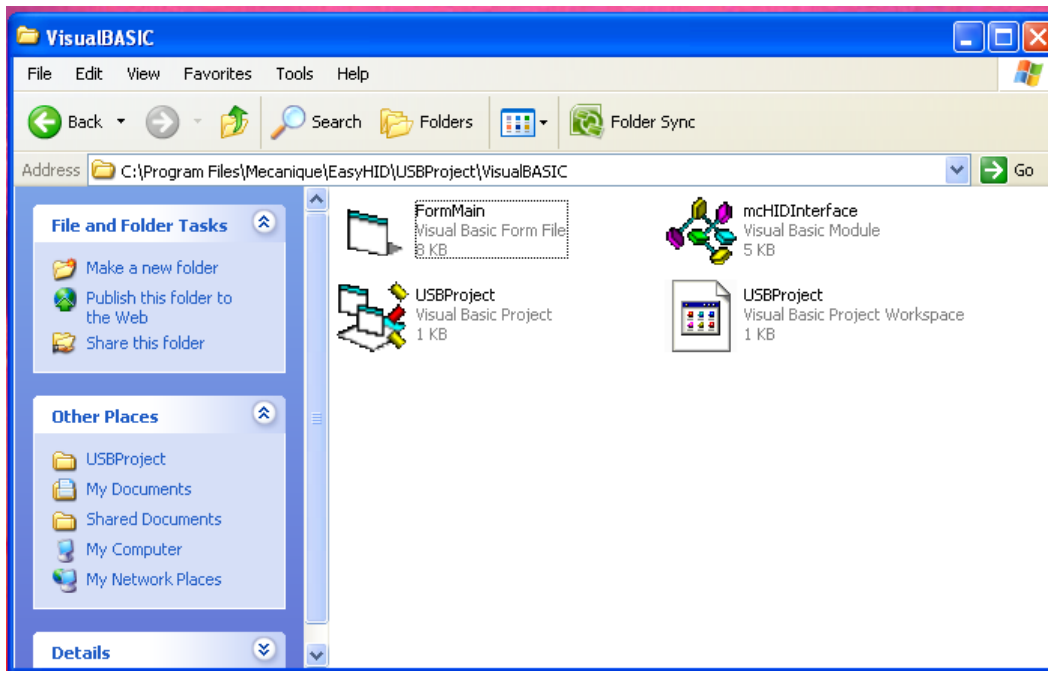

Figure (26): Visual Basic Created Project

## 2.2.2-Microsoft Visual Basic 6.0:

Visual Basic is a programming language which can be use to develop window based applications and games.

There are numbers of the programming languages available for developing applications but the following benefits makes Visual Basic good software to use in this project:

- ✓ It is easy to learn and use programming language compare with others (visual C++, etc).
- ✓ The source availability for learning as it is much more popular compare with others.
- ✓ There are number of tools (shareware & Freeware) on internet that can help to spare some programming time. Such as the Easy hid wizard who was used to create template file for the USB hid.

The template file was created with Easyhid was opened with Visual Basic by simply download and run Visual Basic6.

By selecting menu and open project (file> project> open project) the directory where project was created in previous part by Easy hid was selected and opened.

The opened project has two file (Mainform & HIDDLLInterface). The Mainform was modified and expanded in order to screen the input data as shown in the following Form and has the following controls.

Figure (27) illustrate the Form was displayed when the program was not connected to the device.

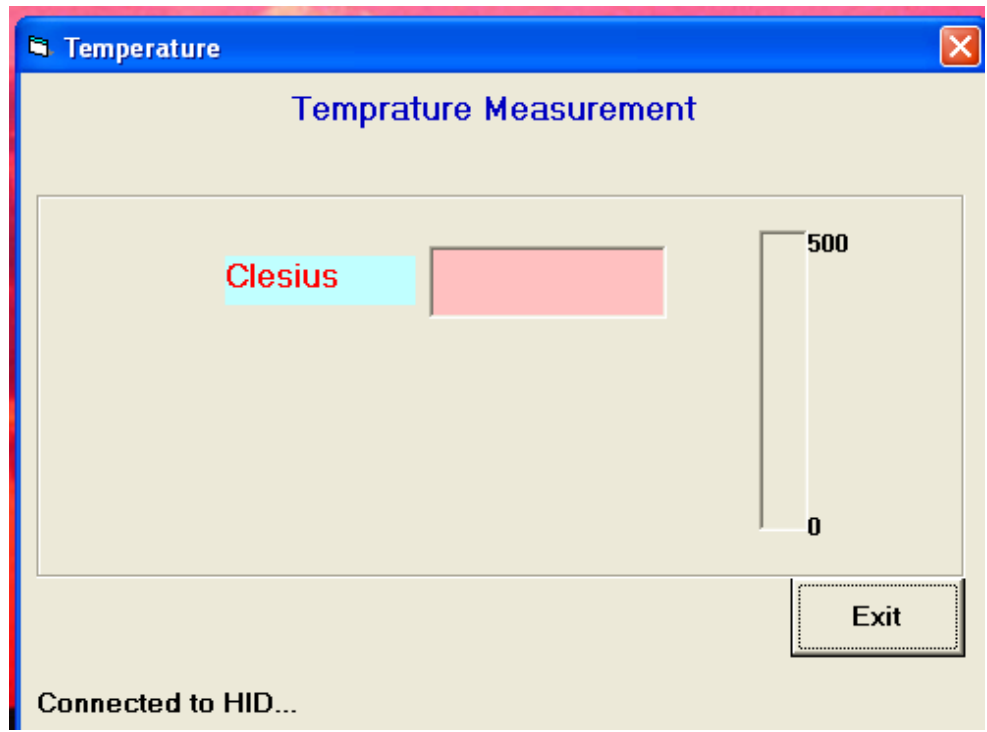

Figure (27): Screen out Form when not is connected to the device

And when the device was connected to the PC using the USB cable the Form was appeared as shown in Figure (28).

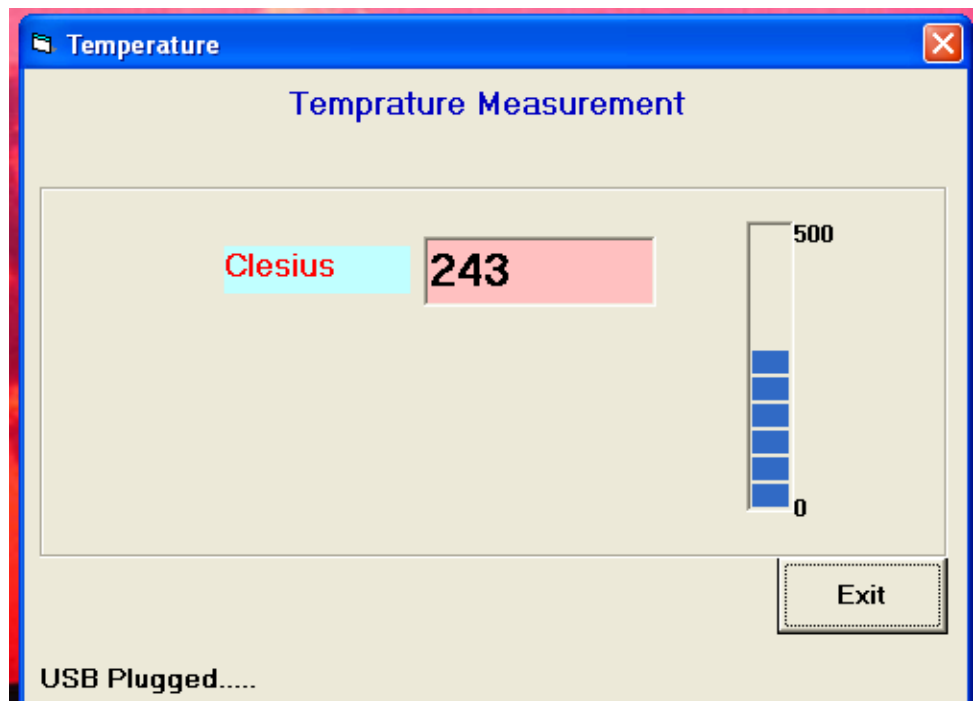

Figure (28): Screen out Form when is connected to the device

In this part of report more details was given from the expansion part of the Mainform file program in order to display above output Form compare with the blank Form was created by the Easy hid.

a) The first part was added to the template file generated by Easy hid was to display messages on the Form that indicated Form was connected to HID and USB plugged, unplugged. This was obtained by dragged a label to the bottom part of Form where the message displays and was renamed to lblstatus and following message was added to the HID connect function in the Mainform coding part of program to get display as this function being executed by the program, when it is running.
lblstatus = "Connected to HID..."
The following messages was added to the USB plugged and unplugged functions in order to display following messages in the case of connection or disconnection of the USB cable as shown here and in the appendix part of report.
lblstatus = "USB Plugged....."
lblstatus = "USB Unplugged....."

Adding these output messages to the end of each operation in the Mainform, will create output message whenever the program executing each specific function.

b) The next and important subroutine has been added the program was the onRead function used to catch data from the USB bus was sent by the microcontroller and display it on the screen.
The subroutine has a global variable (temperature) which stores 4 digits input data and one digit Report ID into 5 buffer inputs is sending from the microcontroller.
The first buffer (BufferIn (0)) always carry Report ID about the packages was sent from the microcontroller.
The next 4 buffers (BufferIn (1), BufferIn (2), BufferIn (3), BufferIn (4)) are the actual 4 bits data sent as a string of character to the PC.

    These characters were stored in string temperature variable and were displayed by textbox (txtno) in next line of coding program.

    Progress bar was another technique was used to output temperature on the bar but as it can output only the integer data the output has to convert from the character to the integer.

    All these operations were created and are shown in OnRead subroutine of Mainform appendix.

c) The next step was to add the command button in the bottom part of Form, the following subroutine (key_click) was added in order to stop and exit from the program.
Form_Unload (0);

d) The labels such as the temperature measurement, Celsius and number were added to the Form to specify what each control does on the Form. By simply add a label and change its name to the name needed and located it next to the control.

e) At the labels and controls was modified to display as it needed by simply changing fonts, size, etc in their properties control page.

The main coding for the Form is shown in appendix3and interface in appendix 4.

## 2.2.3-USBTrace USB protocol analyzer

There are number of USB protocol analyzer in the market that can be use to check data transaction on the USB bus. Some are hardware based which make them more expensive and some are software based that make them cheaper compare with hardware based one.

The USBTrace is one of the software based analyzer that was developed by SysNcleus (www.sysnucleus.com), in this report the USBTrace USB protocol analyzer was used to monitored data on the USB bus in the case of project not operating as it was expected because of any problem. A limited time demo version of USBTrace is available on manufacture's web site.

The following steps shows in this section illustrate processes was taken place in order to use the software to monitor data was sent by the microcontroller to the PC using the USB bus:

- ✓ The first step was to download and install program from manufacture's website.
- ✓ The next step was to double click on the USBTrace icon on the desktop which runs the program and the Figure (29) window will appear on the screen.

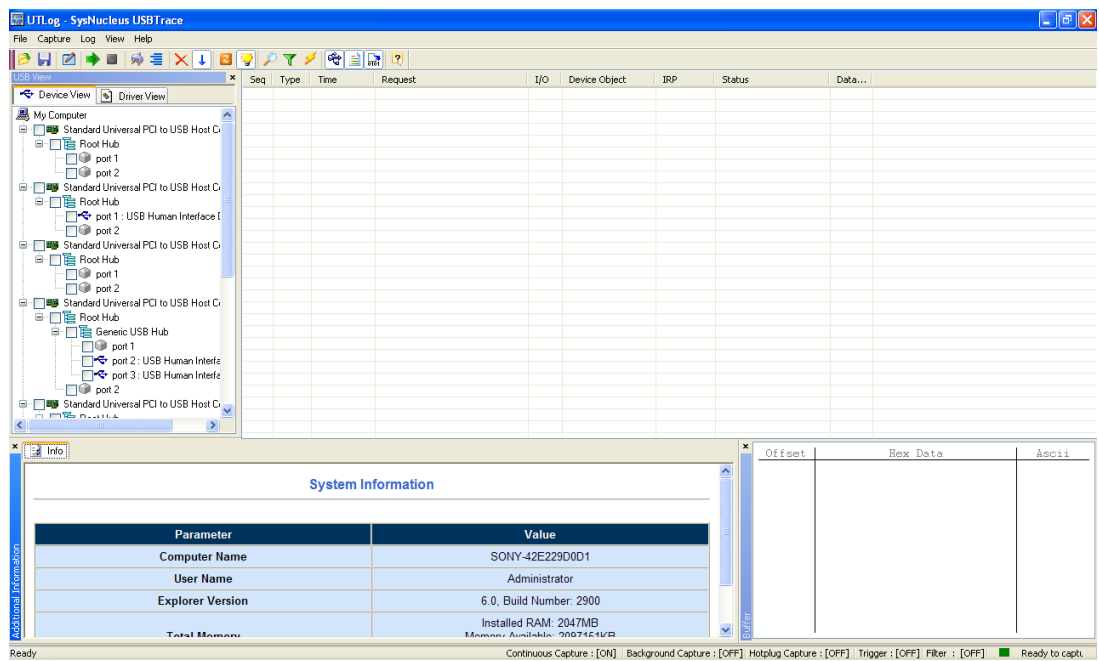

Figure (29): USBTrace Window when the device not connected

- ✓ The microcontroller was connected to the USB port of the PC which was detected by the program and a new human interface device was added to the left part of window.
- ✓ The device was added in the previous step was selected by clicking on the box located next to the HID device.
- ✓ The data is captured by clicking on the green arrow located on the top left of the menu.

The START OF LOG message was appears in the middle of the screen as the button was clicked and data was captured by the program every second as it was expected.

The captured data was indicated by green color this shown in Figure (30).

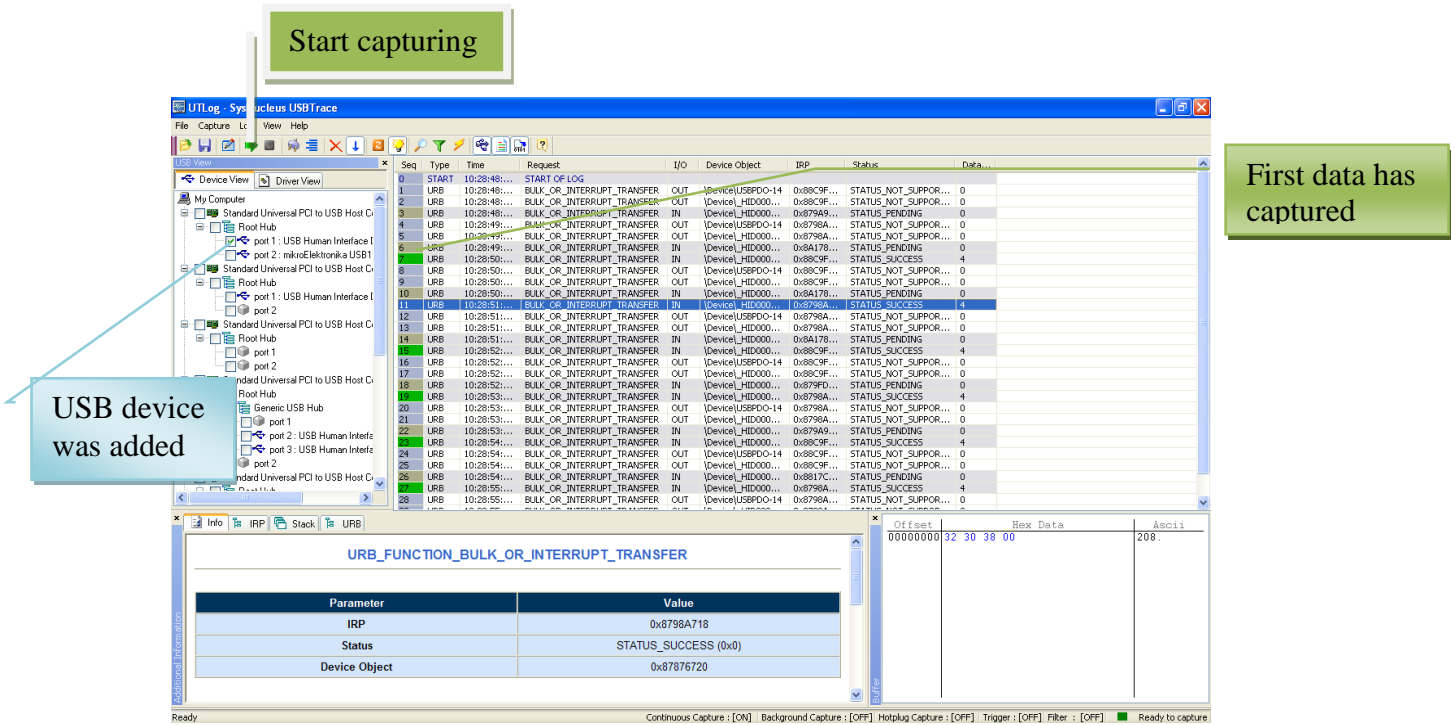

Figure (30): Figure (29): USBTrace Window when the device was connected and ran

- ✓ By moving the curser over each packet, the pop up window will appear that gives more information about packet and data has been send in hexadecimal.

- ✓ By clicking on specific packet, the data value of data in Hexadecimal and ASCII will appear on bottom left of screen as shown in previous figure.

With this software after checking the data on the USB bus and comparing it with the data was expected to screen out on the PC's screen.

The designer can work out where the problem starts by observing and comparing the two data streams. The software can be great help to spot the problem in the design by simply monitoring and checking the data on the bus.

# 5) Results & Discussions:

The project was designed, built and tested. The test results concludes that the software functions as expected from design specification and it is satisfies all requirements.
The test of the project was relatively trouble-free part of the project that can be done by going through following steps:

- ➢ The first step was to construct hardware by placing the PIC18F4550 in the EasyPIC4 development board and activate the part of board is going to be used in the project.

- ➢ The second step was to load the program into PIC18F4550 microcontroller by connecting board to the PC via the USB 2.0 programmer port and load the program using MikroC compiler and PICFlash2 software.

- ➢ Next step was to run the visual basic program that detect human interface device and catch data and output it on the screen.

- ➢ And finally connect the USB connector to the PC that establishes communication and sends data every second. As the microcontroller connects to the PC's port for the first time, a message should pops up on the bottom of screen to indicate the human interface device was recognized by the PC.

  This can be checked through device manager too. The operation indicates the VIP and PID were set to the right values.
  Connecting the USB connector makes Visual Basic program to recognize the device and display out USB Plugged message and start to catching and displaying data on the Form.

The project works as it was expected and it satisfies the customer specification
Every devices and software has been used in this project had advantages over other similar technology on the market, some described in following:

- ✓ The microcontroller PIC18F4550 had 10-bit analog to digital conversion as well as the USB 2.0 technology built in one chip.

- ✓ The EasyPIC4 development board was used as it had all the hardware components needed for the project all on one board built in and was used before for other designed project which made me more familiar to work with its operation compare with other development board.

- ✓ MikroC compiler was used as it was easy to use and had the HID terminal that could generate the USB descriptor file as well as enabling user to communicate with the microcontroller through the USB bus without any other software.

- ✓ Visual Basic software was used as the designer has knowledge of how to using software and it is easier to used and understand compare with other version such as Lab view and Visual C++.

- ✓ Easy HID software was used to generate initial template file for Visual Basic framework which makes design easier.

- ✓ USBTrace protocol analyzer was used as it is software base, easy to use and cheaper compare with other protocol analyzer in the market.

# 6)    Conclusions & Future works:

The designed system can get input analog signal from a temperature sensor and output measured temperature in Celsius on the screen. The output data is updating every second. Using the USB 2.0 technology to send data from sensor to the host computer was one of the main challenges in this project which was involved with interfacing and protocols.

The following changing can be made to the system that can improve the system to be used by any other purposes that involves with sending data using USB 2.0 technology.

- ➢ The potentiometer was used in the project to illustrate the analog signal as a temperature sensor. However in the case of where the sensor was defined, this can be change simply by setting the PORTA as an input and connect the sensor to the port. Choosing the PORTA as an input was done by simply disconnecting JP15 in A/D converter part of EasyPIC4 development board.
- ➢ The system was designed to detect analog signal from the input, but in the case of the digital sensor is going to be used to the input port. The ADCON1 control register has to reconfigure to digital and some part of calculation in the microcontroller coding has to get change.
- ➢ The input sensor can be increase as some device works with more than one sensor by specifying more input channels to read the data inputs and store them in larger write buffer array size that used to hold data on microcontroller part of program as well as expanding Visual Basic coding in such way that enable program to separate inputs data and output them in deferent textboxes on the form.
  Another way to do this is to use multiplexer to switch from one channel to another in each moment of time.
- ➢ The design system can use wireless technology in order to establish communication between the device and the host computer to sending the input data from a sensor to the PC for more operation and store entire data in a database.
- ➢ The data however can be sent using apache software from one place to another for the user attention.

# 7) Project Planning:

This chart deals with planning and management of the project. Planning and management of any project is basically done following the basic structure as to ''WHERE YOU ARE'' and ''WHERE DO YOU WANT TO GO'' and finally giving an idea ''HOW TO GET THERE''.

## Gantt chart:

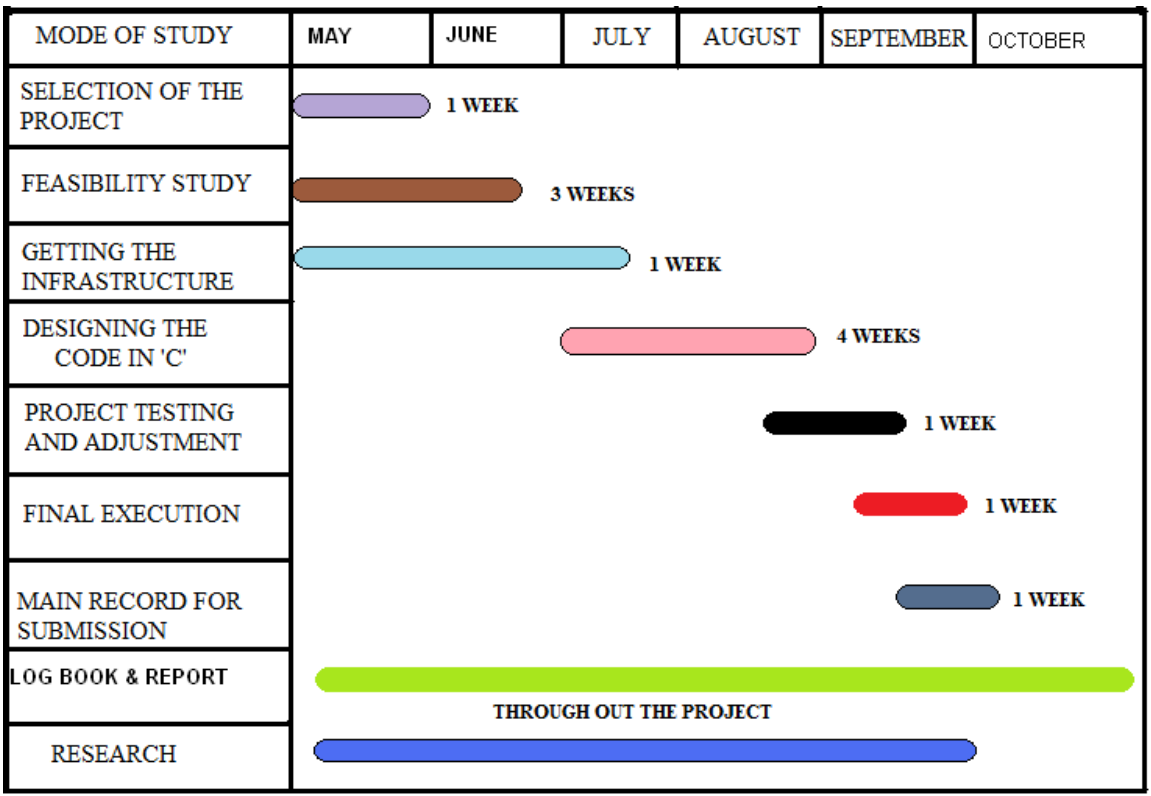

The expected duration of the project is 12 Weeks, which is divided as mentioned above.

**NOTE:** The stipulated timings mentioned to complete the project might change as it is the maximum expected duration, and there is every chance to complete it early.

# 8) References and Appendices:

# 8.2. Appendices:

# Appendix 1: Main program For Microcontroller:

```
/*............................................................
USB BASED SKIN MOISTURE MEASUREMENT
==============================================
In this project a PIC18F4550 type microcontroller is connected to a PC through the USB
cable.
In addition, a Temperature sensor is connected to analog port AN0 of the microcontroller.
The microcontroller reads the sensor and sends it to the PC every second.
The PC displays the reading on the screen.
A Visual Basic program runs on the PC which reads the sensor from the USB port and
then displays it on a form.
The microcontroller is operated from an 8MHz crystal, but the CPU clock frequency is
increased to 48MHz.
Also, the USB module operates with 48MHz.
The Temperature is sent to the PC in Celsius as a 4 digit integer number.
Author: Siamak Sarjoghian
Date: September 2010
File: Temperature.c
==============================================
............................................................*/

unsigned char i,j;
unsigned long Vin, Cint;
unsigned char op[12], Temperature[4], Read_buffer[4];
float Voltage,Celsius;

void System_init(void);
void Interrupt_Dis(void);
void Interrupt_En(void);
void Timer0_init(void);
void Remove_blank(void);
//---------------------------------------------------------
//
// Timer interrupt service routine
//
//---------------------------------------------------------
void interrupt()
{
HID_InterruptProc(); // Keep alive
TMR0L = 100; // Reload TMR0L
INTCON.TMR0IF = 0; // Re-enable TMR0 interrupts
}
```

```c
//---------------------------------------------------------
// System_init(void)
// This function initializes the system.
// Pins used:
// Set inputs as analog, Ref=+5V
// Set A/D clock = Fosc/64, 8TAD
// Set PORT A as inputs
// Set PORT B as outputs
//---------------------------------------------------------
void System_init(void)
{
ADCON1 = 0x00;
ADCON2 = 0xA6;
TRISA = 0xFF;
TRISB = 0X00;
}
//---------------------------------------------------------
// void Interrupt_Dis(void)
//
// Set interrupt registers to power-on defaults
// Disable all interrupts
//
//---------------------------------------------------------
void Interrupt_Dis(void){
INTCON=0x00;
INTCON2=0xF5;
INTCON3=0xC0;
RCON.IPEN=0x00;
PIE1=0x00;
PIE2=0x00;
PIR1=0x00;
PIR2=0x00;
}
//---------------------------------------------------------
// void Interrupt_En(void)
//
// Set interrupt registers to power-on defaults
// Enable all interrupts
//
//---------------------------------------------------------
void Interrupt_En(void){
INTCON = 0xE0;
}
//---------------------------------------------------------
// void Timer0_init(void)
//
```

```c
// Configure TIMER 0 for 3.3ms interrupts. Set prescaler to 256
// and load TMR0L to 156 so that the time interval for timer
// interrupts at 48MHz is 256.156.0.083 = 3.3ms
//
// The timer is in 8-bit mode by default
//
//---------------------------------------------------------
void Timer0_init(void){
T0CON = 0x47; // Prescaler = 256
TMR0L = 100; // Timer count is 256-156 = 100
INTCON.TMR0IE = 1; // Enable T0IE
T0CON.TMR0ON = 1; // Turn Timer 0 ON
}
//---------------------------------------------------------
// Remove_blank(void)
// Remove leading blanks from op array.
// The function save first 5 character in temperature array.
//---------------------------------------------------------
void Remove_blank(void){
for(j=0; j<4; j++)Temperature[j]=' ';
j=0;
for(i=0;i<=11;i++)
{
if(op[i] != ' ') // If a blank
{
Temperature[j]=op[i];
j++;
}
}
}
//---------------------------------------------------------
//
// Start of MAIN program
//
//---------------------------------------------------------
void main()
{
System_init();
Interrupt_Dis();
Timer0_init();
Interrupt_En();
//---------------------------------------------------------
// Enable USB port
//---------------------------------------------------------
Hid_Enable(&Read_buffer, &Temperature);
Delay_ms(1000);// Wait 1 second
```

```c
Delay_ms(1000);// Wait 1 second
//---------------------------------------------------------
//
// Endless loop. Read Temperature from the A/D converter,
// convert into Celsius and send to the PC over the
// USB port every second
// forever loop
//
//---------------------------------------------------------
for(;;)
{
Vin = Adc_Read(0); // Read from channel 0 (AN0)
Voltage = (Vin * 5.0) / 1024.0; // In Voltage = Vin x 5/1024
Celsius = (Voltage * 100.0); // convert voltage to Temperature in Celsius
Cint = (int)Celsius; // As an integer number
PORTB = Cint;
LongToStr(Cint,op); // Convert to string in "op"

Remove_blank();   //To remove blanks characters from op and save it in temperature.
//---------------------------------------------------------
// Send Temperature  (in array Temperature ) to the PC
//---------------------------------------------------------
Hid_Write(&Temperature,4); // Send to USB as 4 characters
Delay_ms(1000); // Wait 1 second
}
//---------------------------------------------------------
// Disable USB port
//---------------------------------------------------------
Hid_Disable();
}
```

# Appendix 2: USB Descriptor File:

```c
//***************************************************************************
//
// File Version 1.01
//
//***************************************************************************

#include "Definit.h"
#include "VARs.h"

//***************************************************************************
// The number of bytes in each report,
// calculated from Report Size and Report Count in the report descriptor
//***************************************************************************
unsigned char const HID_INPUT_REPORT_BYTES    = 4;
unsigned char const HID_OUTPUT_REPORT_BYTES   = 4;

unsigned char const HID_FEATURE_REPORT_BYTES  = 2;
//***************************************************************************
// Byte constants
//***************************************************************************
unsigned char const NUM_ENDPOINTS             = 2;
unsigned char const ConfigDescr_wTotalLength  =
USB_CONFIG_DESCRIPTOR_LEN + USB_INTERF_DESCRIPTOR_LEN +
USB_HID_DESCRIPTOR_LEN + (NUM_ENDPOINTS *
USB_ENDP_DESCRIPTOR_LEN);
unsigned char const HID_ReportDesc_len        = 47;

unsigned char const Low_HID_ReportDesc_len    = HID_ReportDesc_len;
unsigned char const High_HID_ReportDesc_len   = HID_ReportDesc_len >> 8;

unsigned char const Low_HID_PACKET_SIZE       = HID_PACKET_SIZE;
unsigned char const High_HID_PACKET_SIZE      = HID_PACKET_SIZE >> 8;

//***************************************************************************
// Descriptor Tables
//***************************************************************************
```

```c
unsigned char const DescTables[USB_DEVICE_DESCRIPTOR_ALL_LEN*2] = {

// Device Descriptor
    USB_DEVICE_DESCRIPTOR_LEN, 0,     // bLength           - Length of Device descriptor (always 0x12)
    USB_DEVICE_DESCRIPTOR_TYPE, 0,    // bDescriptorType   - 1 = DEVICE descriptor
    0x00, 0,                          // bcdUSB            - USB revision 2.00 (low byte)
    0x02, 0,                          //                     (high byte)
    0x00, 0,                          // bDeviceClass      - Zero means each interface operates independently (class code in the interface descriptor)
    0x00, 0,                          // bDeviceSubClass
    0x00, 0,                          // bDeviceProtocol
    EP0_PACKET_SIZE, 0,               // bMaxPacketSize0   - maximum size of a data packet for a control transfer over EP0
    0x34, 0,                          // idVendor          - Vendor ID (low byte)
    0x12, 0,                          //                     (high byte)
    0x01, 0,                          // idProduct         - Product ID (low byte)
    0x00, 0,                          //                     (high byte)
    0x01, 0,                          // bcdDevice         - ( low byte)
    0x00, 0,                          //                     (high byte)
    0x01, 0,                          // iManufacturer     - String1
    0x02, 0,                          // iProduct          - String2
    0x00, 0,                          // iSerialNumber     - ( None )
    0x01, 0,                          // bNumConfigurations - 1

// Configuration Descriptor
    USB_CONFIG_DESCRIPTOR_LEN, 0,     // bLength           - Length of Configuration descriptor (always 0x09)
    USB_CONFIG_DESCRIPTOR_TYPE, 0,    // bDescriptorType   - 2 = CONFIGURATION descriptor
    ConfigDescr_wTotalLength, 0,      // wTotalLength      - Total length of this config. descriptor plus the interface and endpoint descriptors that are part of the configuration.
    0x00, 0,                          //                     ( high byte)
    0x01, 0,                          // bNumInterfaces    - Number of interfaces
    0x01, 0,                          // bConfigurationValue - Configuration Value
    0x00, 0,                          // iConfiguration    - String Index for this configuration ( None )
    0xA0, 0,                          // bmAttributes      - attributes - "Bus powered" and "Remote wakeup"
    50, 0,                            // MaxPower          - bus-powered draws 50*2 mA from the bus.

// Interface Descriptor
```

```
   USB_INTERF_DESCRIPTOR_LEN, 0,         // bLength            - Length of
Interface descriptor (always 0x09)
   USB_INTERFACE_DESCRIPTOR_TYPE, 0,     // bDescriptorType    - 4 =
INTERFACE descriptor
   0x00, 0,                              // bInterfaceNumber   - Number of interface, 0 based
array
   0x00, 0,                              // bAlternateSetting  - Alternate setting
   NUM_ENDPOINTS, 0,                     // bNumEndPoints      - Number of endpoints
used in this interface
   0x03, 0,                              // bInterfaceClass    - assigned by the USB
   0x00, 0,                              // bInterfaceSubClass - Not A boot device
   0x00, 0,                              // bInterfaceProtocol - none
   0x00, 0,                              // iInterface         - Index to string descriptor that
describes this interface ( None )

// HID Descriptor
   USB_HID_DESCRIPTOR_LEN, 0,            // bLength            - Length of HID
descriptor (always 0x09)
   USB_HID_DESCRIPTOR_TYPE, 0,           // bDescriptorType    - 0x21 = HID
descriptor
   0x01, 0,                              // HID class release number (1.01)
   0x01, 0,
   0x00, 0,                              // Localized country code (none)
   0x01, 0,                              // # of HID class descriptor to follow (1)
   0x22, 0,                              // Report descriptor type (HID)
   Low_HID_ReportDesc_len, 0,
   High_HID_ReportDesc_len, 0,

// EP1_RX Descriptor
   USB_ENDP_DESCRIPTOR_LEN, 0,           // bLength            - length of descriptor
(always 0x07)
   USB_ENDPOINT_DESCRIPTOR_TYPE, 0,      // bDescriptorType    - 5 =
ENDPOINT descriptor
   0x81, 0,                              // bEndpointAddress   - In, EP1
   USB_ENDPOINT_TYPE_INTERRUPT, 0,       // bmAttributes       - Endpoint
Type - Interrupt
   Low_HID_PACKET_SIZE, 0,               // wMaxPacketSize     - max packet size -
low order byte
   High_HID_PACKET_SIZE, 0,              //                    - max packet size - high order
byte
   1, 0,                                 // bInterval          - polling interval (1 ms)

// EP1_TX Descriptor
   USB_ENDP_DESCRIPTOR_LEN, 0,           // bLength            - length of descriptor
(always 0x07)
```

```
    USB_ENDPOINT_DESCRIPTOR_TYPE, 0,    // bDescriptorType    - 5 = ENDPOINT descriptor
    0x01, 0,                            // bEndpointAddress   - Out, EP1
    USB_ENDPOINT_TYPE_INTERRUPT, 0,     // bmAttributes       - Endpoint Type - Interrupt
    Low_HID_PACKET_SIZE, 0,             // wMaxPacketSize     - max packet size - low order byte
    High_HID_PACKET_SIZE, 0,            //                    - max packet size - high order byte
    1, 0,                               // bInterval          - polling interval (1 ms)

// HID_Report Descriptor
    0x06, 0,                    // USAGE_PAGE (Vendor Defined)
    0xA0, 0,
    0xFF, 0,
    0x09, 0,                    // USAGE ID (Vendor Usage 1)
    0x01, 0,
    0xA1, 0,                    // COLLECTION (Application)
    0x01, 0,
//  The Input report
    0x09, 0,                    // USAGE ID - Vendor defined
    0x03, 0,
    0x15, 0,                    //   LOGICAL_MINIMUM (0)
    0x00, 0,
    0x26, 0,                    //   LOGICAL_MAXIMUM (255)
    0x00, 0,
    0xFF, 0,
    0x75, 0,                    //   REPORT_SIZE (8)
    0x08, 0,
    0x95, 0,                    //   REPORT_COUNT (2)
    HID_INPUT_REPORT_BYTEs, 0,
    0x81, 0,                    //   INPUT (Data,Var,Abs)
    0x02, 0,
//  The Output report
    0x09, 0,                    // USAGE ID - Vendor defined
    0x04, 0,
    0x15, 0,                    //   LOGICAL_MINIMUM (0)
    0x00, 0,
    0x26, 0,                    //   LOGICAL_MAXIMUM (255)
    0x00, 0,
    0xFF, 0,
    0x75, 0,                    //   REPORT_SIZE (8)
    0x08, 0,
    0x95, 0,                    //   REPORT_COUNT (2)
    HID_OUTPUT_REPORT_BYTES, 0,
    0x91, 0,                    //   OUTPUT (Data,Var,Abs)
```

```
  0x02, 0,
// The Feature report
  0x09, 0,                      // USAGE ID - Vendor defined
  0x05, 0,
  0x15, 0,                      //  LOGICAL_MINIMUM (0)
  0x00, 0,
  0x26, 0,                      //  LOGICAL_MAXIMUM (255)
  0x00, 0,
  0xFF, 0,
  0x75, 0,                      //  REPORT_SIZE (8)
  0x08, 0,
  0x95, 0,                      //  REPORT_COUNT (2)
  HID_FEATURE_REPORT_BYTES, 0,
  0xB1, 0,                      //  FEATURE (Data,Var,Abs)
  0x02, 0,
// End Collection
  0xC0, 0                       // END_COLLECTION
};
//****************************************************************************
unsigned char const LangIDDescr[8] = {
  0x04, 0,
  USB_STRING_DESCRIPTOR_TYPE, 0,
  0x09, 0,                      // LangID (0x0409) - Low
  0x04, 0                       //               - High
};
//****************************************************************************
unsigned char const ManufacturerDescr[68] = {
  34, 0,
  USB_STRING_DESCRIPTOR_TYPE, 0,
  'm', 0, 0, 0,
  'i', 0, 0, 0,
  'k', 0, 0, 0,
  'r', 0, 0, 0,
  'o', 0, 0, 0,
  'E', 0, 0, 0,
  'l', 0, 0, 0,
  'e', 0, 0, 0,
  'k', 0, 0, 0,
  't', 0, 0, 0,
  'r', 0, 0, 0,
  'o', 0, 0, 0,
  'n', 0, 0, 0,
  'i', 0, 0, 0,
  'k', 0, 0, 0,
```

```c
  'a', 0, 0, 0
};
//****************************************************************
********
unsigned char const ProductDescr[96] = {
  48, 0,
  USB_STRING_DESCRIPTOR_TYPE, 0,
  'm', 0, 0, 0,
  'i', 0, 0, 0,
  'k', 0, 0, 0,
  'r', 0, 0, 0,
  'o', 0, 0, 0,
  'C', 0, 0, 0,
  ' ', 0, 0, 0,
  'H', 0, 0, 0,
  'T', 0, 0, 0,
  'D', 0, 0, 0,
  ' ', 0, 0, 0,
  'L', 0, 0, 0,
  'i', 0, 0, 0,
  'b', 0, 0, 0,
  'r', 0, 0, 0,
  'a', 0, 0, 0,
  'r', 0, 0, 0,
  'y', 0, 0, 0,
  ' ', 0, 0, 0,
  't', 0, 0, 0,
  'e', 0, 0, 0,
  's', 0, 0, 0,
  't', 0, 0, 0
};
//****************************************************************
********
unsigned char const StrUnknownDescr[4] = {
  2, 0,
  USB_STRING_DESCRIPTOR_TYPE, 0
};
//****************************************************************
********
```

```
//****************************************************************************
// Initialization Function
//****************************************************************************
void InitUSBdsc()
{
Byte_tmp_0[0] = NUM_ENDPOINTS;
Byte_tmp_0[0] = ConfigDescr_wTotalLength;
Byte_tmp_0[0] = HID_ReportDesc_len;
Byte_tmp_0[0] = Low_HID_ReportDesc_len;
Byte_tmp_0[0] = High_HID_ReportDesc_len;
Byte_tmp_0[0] = Low_HID_PACKET_SIZE;
Byte_tmp_0[0] = High_HID_PACKET_SIZE;

DescTables;

LangIDDescr;
ManufacturerDescr;
ProductDescr;
StrUnknownDescr;

}
//****************************************************************************
```

# Appendix 3: Visual Basic Main Form Code:

```vb
Private Const VendorID = 4660
Private Const ProductID = 1

' read and write buffers
Private Const BufferInSize = 8
Private Const BufferOutSize = 8
Dim BufferIn(0 To BufferInSize) As Byte
Dim BufferOut(0 To BufferOutSize) As Byte
' *****************************************************************
' when the form loads, connect to the HID controller - pass
' the form window handle so that you can receive notification
' events...
'******************************************************************
Private Sub Form_Load()
  ConnectToHID (Me.hwnd)
  lblstatus = "Connected to HID..."
End Sub

'******************************************************************
' disconnect from the HID controller...
'******************************************************************
Private Sub Form_Unload(Cancel As Integer)
  DisconnectFromHID
End Sub

'******************************************************************
' A HID device has been plugged in...
'******************************************************************
Public Sub OnPlugged(ByVal pHandle As Long)
  If hidGetVendorID(pHandle) = VendorID And hidGetProductID(pHandle) = ProductID Then
  lblstatus = "USB Plugged....."
  End If

End Sub

'******************************************************************
' A HID device has been unplugged...
'******************************************************************
Public Sub OnUnplugged(ByVal pHandle As Long)
  If hidGetVendorID(pHandle) = VendorID And hidGetProductID(pHandle) = ProductID Then
  lblstatus = "USB Unplugged...."
```

```vb
    End If
End Sub

'*******************************************************************
' controller changed notification - called
' after ALL HID devices are plugged or unplugged
'*******************************************************************
Public Sub OnChanged()
  Dim DeviceHandle As Long

  DeviceHandle = hidGetHandle(VendorID, ProductID)
  hidSetReadNotify DeviceHandle, True
End Sub

'*******************************************************************
' on read event...
'*******************************************************************
Public Sub OnRead(ByVal pHandle As Long)
Dim Temperature As String
If hidRead(pHandle, BufferIn(0)) Then

Temperature = Chr(BufferIn(1)) & Chr(BufferIn(2)) & Chr(BufferIn(3)) & Chr(BufferIn(4))
txtno = Temperature
' Display.Text = txtno
  End If
End Sub
'*******************************************************************
' button key to end program
'*******************************************************************
Private Sub Key_Click()
Form_Unload (0)
End
End Sub
```

# Appendix 4: HID Interface API Declaration File:

' this is the interface to the HID controller DLL - you should not
' normally need to change anything in this file.
'
' WinProc() calls your main form 'event' procedures - these are currently
' set to..
'
' MainForm.OnPlugged(ByVal pHandle as long)
' MainForm.OnUnplugged(ByVal pHandle as long)
' MainForm.OnChanged()
' MainForm.OnRead(ByVal pHandle as long)

Option Explicit

' HID interface API declarations...
Declare Function hidConnect Lib "mcHID.dll" Alias "Connect" (ByVal pHostWin As Long) As Boolean
Declare Function hidDisconnect Lib "mcHID.dll" Alias "Disconnect" () As Boolean
Declare Function hidGetItem Lib "mcHID.dll" Alias "GetItem" (ByVal pIndex As Long) As Long
Declare Function hidGetItemCount Lib "mcHID.dll" Alias "GetItemCount" () As Long
Declare Function hidRead Lib "mcHID.dll" Alias "Read" (ByVal pHandle As Long, ByRef pData As Byte) As Boolean
Declare Function hidWrite Lib "mcHID.dll" Alias "Write" (ByVal pHandle As Long, ByRef pData As Byte) As Boolean
Declare Function hidReadEx Lib "mcHID.dll" Alias "ReadEx" (ByVal pVendorID As Long, ByVal pProductID As Long, ByRef pData As Byte) As Boolean
Declare Function hidWriteEx Lib "mcHID.dll" Alias "WriteEx" (ByVal pVendorID As Long, ByVal pProductID As Long, ByRef pData As Byte) As Boolean
Declare Function hidGetHandle Lib "mcHID.dll" Alias "GetHandle" (ByVal pVendoID As Long, ByVal pProductID As Long) As Long
Declare Function hidGetVendorID Lib "mcHID.dll" Alias "GetVendorID" (ByVal pHandle As Long) As Long
Declare Function hidGetProductID Lib "mcHID.dll" Alias "GetProductID" (ByVal pHandle As Long) As Long
Declare Function hidGetVersion Lib "mcHID.dll" Alias "GetVersion" (ByVal pHandle As Long) As Long
Declare Function hidGetVendorName Lib "mcHID.dll" Alias "GetVendorName" (ByVal pHandle As Long, ByVal pText As String, ByVal pLen As Long) As Long
Declare Function hidGetProductName Lib "mcHID.dll" Alias "GetProductName" (ByVal pHandle As Long, ByVal pText As String, ByVal pLen As Long) As Long
Declare Function hidGetSerialNumber Lib "mcHID.dll" Alias "GetSerialNumber" (ByVal pHandle As Long, ByVal pText As String, ByVal pLen As Long) As Long
Declare Function hidGetInputReportLength Lib "mcHID.dll" Alias "GetInputReportLength" (ByVal pHandle As Long) As Long

```vb
Declare Function hidGetOutputReportLength Lib "mcHID.dll" Alias
"GetOutputReportLength" (ByVal pHandle As Long) As Long
Declare Sub hidSetReadNotify Lib "mcHID.dll" Alias "SetReadNotify" (ByVal pHandle
As Long, ByVal pValue As Boolean)
Declare Function hidIsReadNotifyEnabled Lib "mcHID.dll" Alias
"IsReadNotifyEnabled" (ByVal pHandle As Long) As Boolean
Declare Function hidIsAvailable Lib "mcHID.dll" Alias "IsAvailable" (ByVal
pVendorID As Long, ByVal pProductID As Long) As Boolean

' windows API declarations - used to set up messaging...
Private Declare Function CallWindowProc Lib "user32" Alias "CallWindowProcA"
(ByVal lpPrevWndFunc As Long, ByVal hwnd As Long, ByVal Msg As Long, ByVal
wParam As Long, ByVal lParam As Long) As Long
Private Declare Function SetWindowLong Lib "user32" Alias "SetWindowLongA"
(ByVal hwnd As Long, ByVal nIndex As Long, ByVal dwNewLong As Long) As Long

' windows API Constants
Private Const WM_APP = 32768
Private Const GWL_WNDPROC = -4

' HID message constants
Private Const WM_HID_EVENT = WM_APP + 200
Private Const NOTIFY_PLUGGED = 1
Private Const NOTIFY_UNPLUGGED = 2
Private Const NOTIFY_CHANGED = 3
Private Const NOTIFY_READ = 4

' local variables
Private FPrevWinProc As Long    ' Handle to previous window procedure
Private FWinHandle As Long      ' Handle to message window

' Set up a windows hook to receive notification
' messages from the HID controller DLL - then connect
' to the controller
Public Function ConnectToHID(ByVal pHostWin As Long) As Boolean
  FWinHandle = pHostWin
  ConnectToHID = hidConnect(FWinHandle)
  FPrevWinProc = SetWindowLong(FWinHandle, GWL_WNDPROC, AddressOf
WinProc)
End Function

' Unhook from the HID controller and disconnect...
Public Function DisconnectFromHID() As Boolean
  DisconnectFromHID = hidDisconnect
  SetWindowLong FWinHandle, GWL_WNDPROC, FPrevWinProc
End Function
```

```vb
' This is the procedure that intercepts the HID controller messages...
Private Function WinProc(ByVal pHWnd As Long, ByVal pMsg As Long, ByVal wParam As Long, ByVal lParam As Long) As Long
  If pMsg = WM_HID_EVENT Then
    Select Case wParam

      ' HID device has been plugged message...
      Case Is = NOTIFY_PLUGGED
        MainForm.OnPlugged (lParam)

      ' HID device has been unplugged
      Case Is = NOTIFY_UNPLUGGED
        MainForm.OnUnplugged (lParam)

      ' controller has changed...
      Case Is = NOTIFY_CHANGED
        MainForm.OnChanged

      ' read event...
      Case Is = NOTIFY_READ
        MainForm.OnRead (lParam)
    End Select

  End If

  ' next...
  WinProc = CallWindowProc(FPrevWinProc, pHWnd, pMsg, wParam, lParam)

End Function
```